\documentclass[aps,twocolumn,superscriptaddress]{revtex4}
\usepackage{graphicx}

\begin{document}
\title{Nuclear pasta structures and the charge screening effect}

\author{Toshiki Maruyama}
\email{maru@hadron02.tokai.jaeri.go.jp}
\affiliation{Advanced Science Research Center, Japan Atomic Energy Research Institute, Tokai, Ibaraki 319-1195, Japan}
\author{Toshitaka Tatsumi}
\email{tatsumi@ruby.scphys.kyoto-u.ac.jp}
\affiliation{Department pf Physics, Kyoto University, Kyoto, 606-8502, Japan}
\author{Dmitri~N.~Voskresensky}
\email{D.Voskresensky@gsi.de}
\affiliation{Moscow Institute for Physics and Engineering, Kashirskoe sh.~31, Moscow 115409, Russia}
\affiliation{Gesellschaft
 f\"ur Schwerionenforschung mbH, Planckstr. 1,
64291 Darmstadt, Germany}
\author{Tomonori Tanigawa}
\email{tanigawa@tiger02.tokai.jaeri.go.jp}
\affiliation{Japan Society for the Promotion of Science, Tokyo 102-8471, Japan}
\affiliation{Advanced Science Research Center, Japan Atomic Energy Research Institute, Tokai, Ibaraki 319-1195, Japan}
\author{Satoshi Chiba}
\email{sachiba@popsvr.tokai.jaeri.go.jp}
\affiliation{Advanced Science Research Center, Japan Atomic Energy Research Institute, Tokai, Ibaraki 319-1195, Japan}

\date{\today}

\begin{abstract}
Non uniform structures of the nucleon matter at subnuclear densities are
numerically studied by means of the density functional theory with
relativistic mean-fields coupled with  the electric field.
A particular role of the charge
screening effects is demonstrated.
\end{abstract}

\maketitle

\section{Introduction}

There emerged many studies of the mixed phases at various first
order phase transitions such as hadron-quark
deconfinement transition\cite{gle92,HPS93,pei95,voskre,voskre1,emaru1},
kaon condensation \cite{GS99,CG00,CGS00,PREPL00,MYTT,RB,NR00,maruKaon,dubna},
color superconductivity \cite{ARRW,bed,RR04}, superfluidity in atomic traps
\cite{BCR03}, nuclear pasta
\cite{Rav83,Has84,Wil85,Oya93,Lor93,Cheng97,Mar98,kido00,Gen00,Gen02,Gen03}, etc.

At very low densities, nuclei in matter are expected to form the
Coulomb lattice embedded in the neutron-electron seas, that
minimizes the Coulomb interaction energy. With an increase of 
the density, ``nuclear pasta'' structures emerge \cite{Rav83}:
stable nuclear shape may change from droplet to rod, slab,
tube, and to bubble. Pastas are eventually
dissolved into uniform matter at a certain nucleon density below
the saturation density, $\rho_0\simeq 0.16~$fm$^{-3}$. Existence
of pasta phases instead of the separated crystalline lattice of
nuclei and the liquid $npe$ phase would modify some important
processes by changing the hydrodynamic properties and the neutrino
opacity in the supernova matter and in the matter of newly born
neutron stars \cite{horowitz}. Also the pasta phases may influence
neutron star quakes and pulsar glitches via the change of
mechanical properties of the crust matter \cite{MI95}.

A number of authors have investigated the low-density nuclear
matter using various models
\cite{Rav83,Has84,Wil85,Oya93,Lor93,Cheng97,Mar98,kido00,Gen00,Gen02}.
Roughly speaking, the favorable nuclear shape is determined by a
balance between the surface and Coulomb energies. In most of
the previous studies the rearrangement effect of the density
profile of the charged particles due to the Coulomb interaction is
discarded. In Ref.\ \cite{Gen03} the electron screening effect has
been studied and it has been found that this effect is of
minor importance. However, the rearrangement of the proton
profiles as the consequence of the Coulomb repulsion was not shown
up in their model.

A naive application  of Gibbs conditions to separate bulk phases  at the first order
phase transitions, when one ignores
the surface and Coulomb interaction, demonstrates a broad region of
the structured mixed phase,  cf.\ \cite{gle92,CG00}.
However the charge screening effect (caused by the non-uniform charged particle distributions)
should be very important when the typical structure size is of the order of
the minimal Debye screening length in the problem.
It may
largely  affect the stability condition of the geometrical structures in the
mixed phases.
We have been recently exploring the effect of the charge screening in the
context of the various structured mixed phases
\cite{voskre,voskre1,emaru1,maruKaon,dubna,maru1}.
In fact, we have examined the mixed phase at the
quark-hadron transition, kaon condensation and of nuclear pasta, and found that
in cases of the
quark-hadron transition and kaon condensation
the mixed phase might be largely limited by the charge screening
and surface effects.

Our purpose here is, following our preliminary study \cite{dubna,maru1}, to 
investigate the nuclear pasta structures by means of
a relativistic mean-field (RMF) model, which on the one hand does not need an introduction of
the surface tension and on the other hand includes the Coulomb interaction in a proper way.
We figure out how the charge screening effects modify
the results obtained disregarding these effects.
In Sec.\ \ref{Func} we formulate the model and describe our numerical procedure.
In Sec.\ \ref{Bulk} we demonstrate the efficiency of the model in the description of properties of finite nuclei.
Then in Sec.\ \ref{Non-uniform} we describe non-uniform pasta structures first at fixed 
proton to baryon number ratio
that may have an application to the  supernova matter and to  the matter
of a newly born hot protoneutron star.
Then we investigate nuclear pasta at the beta-equilibrium, as they occur in cold neutron stars.
In Sec.\ \ref{CompCalc} we elucidate the effects of the surface and the charge screening.
Then in Sec.\ \ref{Sum} we arrive at the conclusions.

\section{Density functional theory with relativistic mean field}\label{Func}

\subsection{Thermodynamic potential and equations of motion}

Following the idea of the density functional theory (DFT) with
the RMF model \cite{refDFT}, we can formulate
equations of motion to  study non-uniform nuclear matter numerically.
The RMF model with fields of mesons and baryons introduced
in a Lorentz-invariant way is rather simple for numerical calculations,
but realistic enough to reproduce the bulk properties of nuclear matter.
In our framework, the Coulomb interaction is properly included in
equations of motion for nucleons, electrons, and meson mean-fields,
and we solve the Poisson equation for the Coulomb potential $V_{\rm Coul}$
self-consistently with them.
Thus the baryon and electron density profiles, as well as the meson
mean-fields, are determined in a way fully
consistent with the Coulomb potential.

Note that our framework can be easily extended to other situations;
for example, if we take into account kaon either pion condensations,
which are likely realized in a high-density region, we should only add
the relevant  meson field terms.
In Ref.\ \cite{maruKaon} we have included the kaon degree of freedom
in such a treatment to discuss kaon condensation in high density regime.

To begin with, we present the thermodynamic potential for
the system of neutrons, protons and electrons with
 chemical potentials, $\mu_a$ ($a=n,p,e$), respectively;
\begin{equation}\label{Omega-tot}
\Omega\!=\!\Omega_N+\Omega_M
    +\Omega_e,
\end{equation}
where
\begin{equation}\label{eq:OmegaN}
\Omega_N \!=\!
  \sum_{a=p,n}
  \int\!\!\! d^3r\!
  \left[
  \int_0^{k_{{\rm F},a}}
  { d^3k \over 4\pi^3}
  \sqrt{{m_N^*}^2+k^2}-\rho_a\nu_a
  \right]
  ,
\end{equation}
with the local Fermi momenta, $k_{{\rm F},a}({\bf r})$ ($a=n,p$), for nucleons,
\begin{eqnarray}
&&\Omega_M\!=\!\int\!\!\! d^3r\!\left[\!
  {(\nabla\sigma)^2\!+\!m_\sigma^2\sigma^2 \over2}\!+\!U(\sigma)
  \right.\nonumber\\
  &&\left.\ \ \ \ \ \
  -{(\nabla\omega_0)^2\!+\!m_\omega^2\omega_0^2 \over2}
  -{(\nabla R_0)^2\!+\!m_\rho^2R_0^2\over2} \!\right]\!\!,
 \label{eq:OmegaM}
\end{eqnarray}
for the scalar ($\sigma$) and vector mean-fields ($\omega_0, R_0$) and
\begin{equation}
\Omega_e\!=\!\int\!\!\! d^3r\!\left[
-{1\over8\pi e^2}(\nabla {V_{\rm Coul}})^2-{(\mu_e-{V_{\rm Coul}})^4\over12\pi^2}
\right],
\end{equation}
for electrons and the Coulomb potential, $V_{\rm Coul}({\bf r})$, where
$\nu_p({\bf r})=\mu_p+{V_{\rm Coul}({\bf r})}-g_{\omega N}\omega_0({\bf
r})-g_{\rho N}R_0({\bf r}),\ \
\nu_n({\bf r})=\mu_n-g_{\omega N}\omega_0({\bf r})+g_{\rho N}R_0({\bf r}),\ \
m_N^*({\bf r})=m_N-g_{\sigma N}\sigma({\bf r})$,
and the nonlinear potential for the scalar field,
$U(\sigma)={1\over3}bm_N(g_{\sigma N}\sigma)^3+{1\over4}c(g_{\sigma N}\sigma)^4$.
Temperature is assumed to be zero in the present study.

Here we use the local-density approximation for nucleons and electrons.
Strictly speaking, the introduction of the density variable is meaningful,
if the typical length of the nucleon density variation  inside the structure
is larger than the inter-nucleon distance.
We must also bear in mind that
for small structure sizes, quantum effects become prominent which we disregarded.
For the sake of simplicity we also discard nucleon and electron
density derivative terms \cite{refDFT}.
In the case when we suppress derivative terms of nucleon densities
they follow changes of the other  fields that have derivative terms. In our
case these fields are the meson mean-fields and the Coulomb field.
Here we consider large-size pasta structures and
simply discard the density variation effect, as a first-step calculation,
while it can be easily incorporated in the quasi-classical
manner by  the derivative expansion within
the density functional theory \cite{refDFT}.
We also could use the fact that the resulting Debye screening lengths of electrons
and protons characterizing  the
Coulomb field profile are typically much larger than those for all meson mean-fields.
Then we could reduce contribution of the latter to the surface tension term.
%
If the nucleon (neutron and proton) length scales were shorter than
those of changes of the meson mean-fields, one could
simplify the problem by dropping them and
introducing instead a
surface tension term.
This simplified treatment is discussed in detail elsewhere \cite{voskresurf}.
In this paper we avoid this simplification and solve the coupled-channel
problem for the meson mean-fields and the Coulomb field numerically.
Parameters of the RMF model are set to reproduce saturation properties
of nuclear matter:
the minimum energy per baryon $-16.3$ MeV at $\rho =\rho_0 =0.153$ fm$^{-3}$,
the incompressibility $K(\rho_0) =240$ MeV, the effective nucleon  mass
$m_N^{*}(\rho_0)=0.78m_N$; $m_N =938$~ MeV, and the symmetry energy coefficient $a_{\rm sym}=32.5$~ MeV.
Coupling constants and meson masses used in our calculation are listed in Table I.
\begin{table*}
\caption{
Parameter set used in RMF in our calculation.
}
\begin{ruledtabular}
\begin{tabular}{cccccccc}
$g_{\sigma N}$ & 
$g_{\omega N}$ &
$g_{\rho N}$ &
$b$ &
$c$ &
$m_\sigma$ [MeV]&
$m_\omega$ [MeV]&
$m_\rho$ [MeV]\\
\hline\\
6.3935 & 
8.7207 & 
4.2696 & 
0.008659 &
0.002421 &
 400 &
 783 &
 769
\end{tabular}
\end{ruledtabular}
\end{table*}

 From the variational principle
${\delta\Omega\over\delta\phi_i({\bf r})}=0$
($\phi_i=\sigma,R_0,\omega_0,V_{\rm Coul}$) and
${\delta\Omega\over\delta\rho_a({\bf r})}=0$ ($a=n,p,e$),
we get the coupled equations of motion for the mean-fields and the Coulomb
potential,
\begin{eqnarray}
\nabla^2\sigma({\bf r}) &=& m_\sigma^2\sigma({\bf r}) +{dU\over d\sigma}
\nonumber\\ &&
\quad -g_{\sigma N}(\rho_n^{(s)}({\bf r})+\rho_p^{(s)}({\bf r}))
    , \label{sigm}\\
\nabla^2\omega_0({\bf r}) &=& m_\omega^2\omega_0({\bf r}) -g_{\omega N}(\rho_p({\bf r})+\rho_n({\bf r}))
    , \label{omeg}\\
\nabla^2R_0({\bf r}) &=& m_\rho^2R_0({\bf r}) -g_{\rho N}(\rho_p({\bf r})-\rho_n({\bf r}))
    ,\\
\nabla^2{V_{\rm Coul}({\bf r})} &=& 4\pi e^2{\rho_{\rm ch}({\bf r})},  \label{puas}\ \
\end{eqnarray}
with the scalar densities $\rho_a^{(s)}({\bf r})$ ($a=n,p$), 
and the charge density, ${\rho_{\rm ch}({\bf r})}={\rho_p({\bf
r})}+{\rho_e({\bf r})}$. Equations
of motion for fermions yield the standard relations between the densities
and chemical potentials,
\begin{eqnarray}
\mu_n &=&\mu_B
   =  \sqrt{k_{{\rm F},n}^2({\bf r})+{m_N^*({\bf r})}^2}\nonumber\\
&&\quad +g_{\omega N}\omega_0({\bf r})-g_{\rho N}R_0({\bf r}),
  \label{eq:cpotB}\\
\mu_p &=&\mu_B-\mu_e
   =  \sqrt{k_{{\rm F},p}^2({\bf r})+{m_N^*({\bf r})}^2}\nonumber\\
&&\quad +g_{\omega N}\omega_0({\bf r})+
  g_{\rho N}R_0({\bf r})-{V_{\rm Coul}({\bf r})}\label{eq:cpotBp}, \\
{\rho_e({\bf r})}&=&-(\mu_e-{V_{\rm Coul}({\bf r})})^3/3\pi^2,
  \label{eq:rhoe}
\end{eqnarray}
where we have assumed that the system is in chemical equilibrium among nucleons and electrons
and introduced the baryon-number chemical potential $\mu_B$
and the electron-number chemical potential $\mu_e$.
Note that first, the Poisson equation for the Coulomb field (\ref{puas})
is highly nonlinear in
$V_{\rm Coul}({\bf r})$, since  $\rho_{\rm ch}({\bf r})$ in r.h.s.\ includes it in a
complicated way.  The Coulomb potential always enters
equations through the gauge invariant combinations $\mu_e-V_{\rm
Coul}({\bf r})$ and $\mu_p +V_{\rm
Coul}({\bf r})$.


\subsection{Numerical procedure}

To solve the above coupled equations numerically,
we use the Wigner-Seitz cell approximation:
the whole space is divided into equivalent cells with a geometry.
The geometrical shape of the cell changes:
sphere in three-dimensional (3D) calculation, cylinder in 2D and slab in 1D,
respectively.
Each cell is globally charge-neutral and all the  physical quantities
in a cell are smoothly connected to those of the next cell
with zero gradients at the boundary.
Every point inside the cell is represented by the grid points ($N_{\rm grid}\approx 100$) and
the differential equations for fields are solved by the relaxation method
for a given baryon-number density under the constraints of the global charge
neutrality.

To illustrate how to numerically solve equations of motion for the mean-fields,
let us consider, for simplicity, two fields $f_1(r)$, $f_2(r)$
and their coupled Poisson-like equations under 3D calculation,
\begin{eqnarray}
\nabla^2f_1(r)&=&{m_1}^2f_1(r)+W_1[f_1,f_2],\nonumber\\
\nabla^2f_2(r)&=&{m_2}^2f_2(r)+W_2[f_1,f_2],
\label{eq:poisson}
\end{eqnarray}
where $W_i$ ($i=1,2$) are functions of the fields $f_1$ and $f_2$.
Introducing a relaxation ``time'' $t$ artificially, we solve the equation,
\begin{equation}
{\partial f_i(r;t)\over \partial t} = c_i\left(\nabla^2f_i(r;t)-m^2f_i(r;t)-W_i[f_1,f_2]\right).
\label{eq:solvemeson}
\end{equation}
If the coefficients $c_i$ are appropriately chosen, the above $f_i(r;t)$ will
converge to be constant in time and we get the solution of Eq.~(\ref{eq:poisson}).

The profiles of the nucleon densities are solved
with the help of the ``local chemical potentials'' $\mu_a(r)$ ($a=n,p$),
being different from the constant  chemical potentials which we have
initially introduced.
Assuming
$\mu_a(r)$
being an increasing function of the neutron or proton number density
$\rho_a(r)$ in Eqs.~(\ref{eq:cpotB}) and (\ref{eq:cpotBp}),
the relaxation equation for the neutron or proton density profile,
\begin{equation}
{\partial\rho_a(r;t)\over \partial t} = c_a(r;t)\; \rho_a(r;t)\nabla^2\mu_a(r;t),
\label{eq:solvebaryon}
\end{equation}
is solved to get rid of the spatial dependence of
the local chemical potentials $\mu_a(r;t)$. 
%
%
%
The coefficients $c_a(r;t)$ ($a=n,p$) are not constant
so as to conserve the
total proton and neutron numbers. 
When we impose the beta-equilibrium
condition, proton and neutron densities are adjusted to achieve
$\mu_n(r)=\mu_p(r)+\mu_e(r)$. 
Finally we get the density profiles $\rho_n(r)$ and $\rho_p(r)$ relating to the constant
chemical potentials $\mu_n(r)=\mu_n$ and $\mu_p(r)=\mu_p$.
Although the basic idea is to attain the constant chemical
potentials, $\mu_a(r)=\mu_a$ ($a=n,p$) at the convergence, there
is an exception: when there are some regions where $\rho_a(r)=0$,
the local chemical potentials $\mu_a(r)$ are larger than the
constant value in the regions where 
$\rho_a(r)\neq 0$.

The electron density profile $\rho_e(r)$ is calculated directly from
Eq.~(\ref{eq:rhoe}).
The value of $\mu_e$ is adjusted at any time step
to maintain the global charge neutrality:
we decrease $\mu_e$ when the total charge in a cell is positive and
increase when it is negative.

All the above relaxation procedures are performed simultaneously.

\section{Bulk properties of finite nuclei}\label{Bulk}

Before applying our framework to the problem of the pasta phase
in nucleon matter, we check how it works to describe finite nuclei.
In this calculation, for simplicity, we assume  the spherical shape of nuclei.
The electron density is set to be zero. Therefore neither
the global charge neutrality
condition nor the local charge-neutrality condition is imposed.

\begin{figure}
  \includegraphics[width=.23\textwidth]{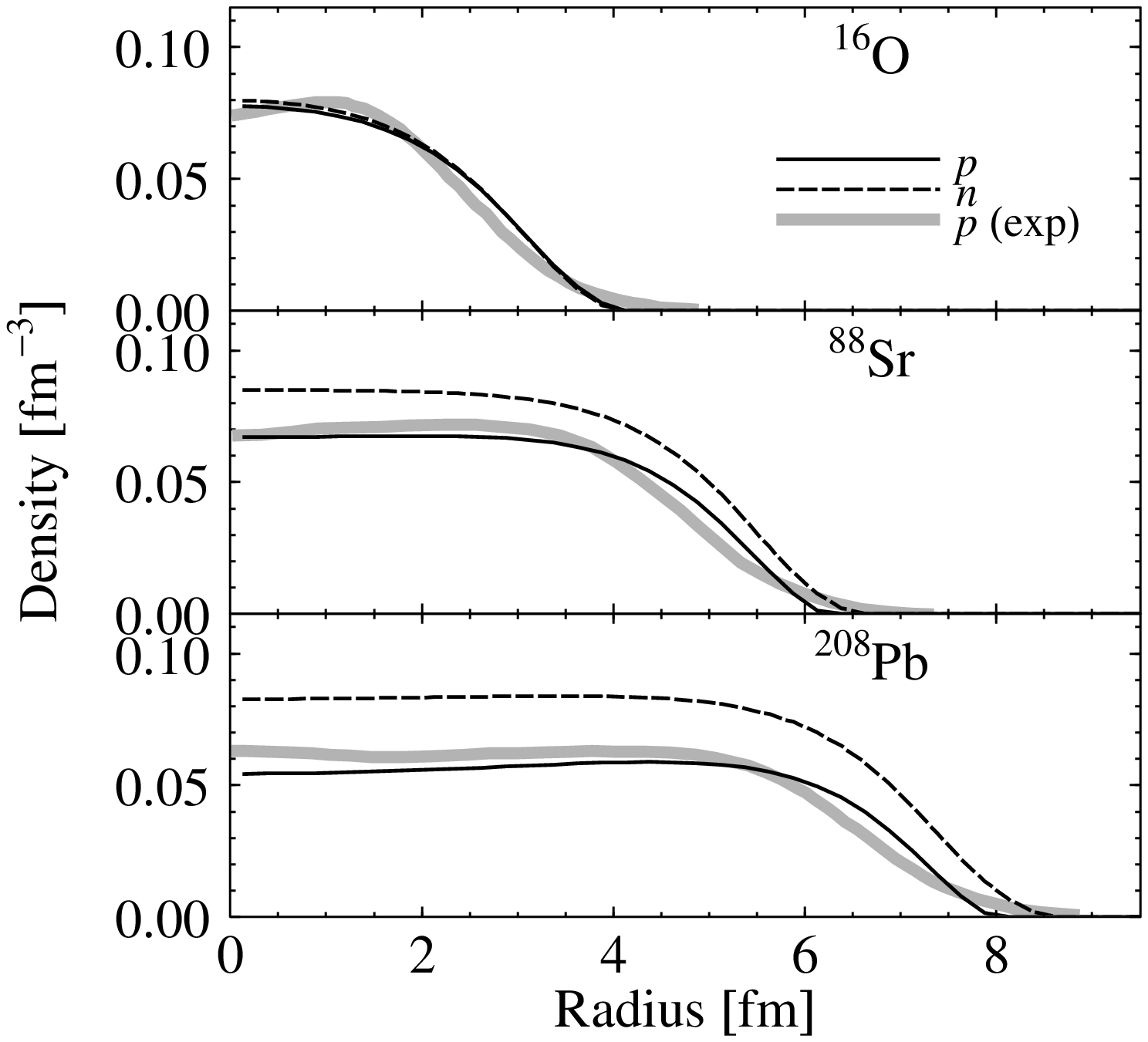}
  \includegraphics[width=.24\textwidth]{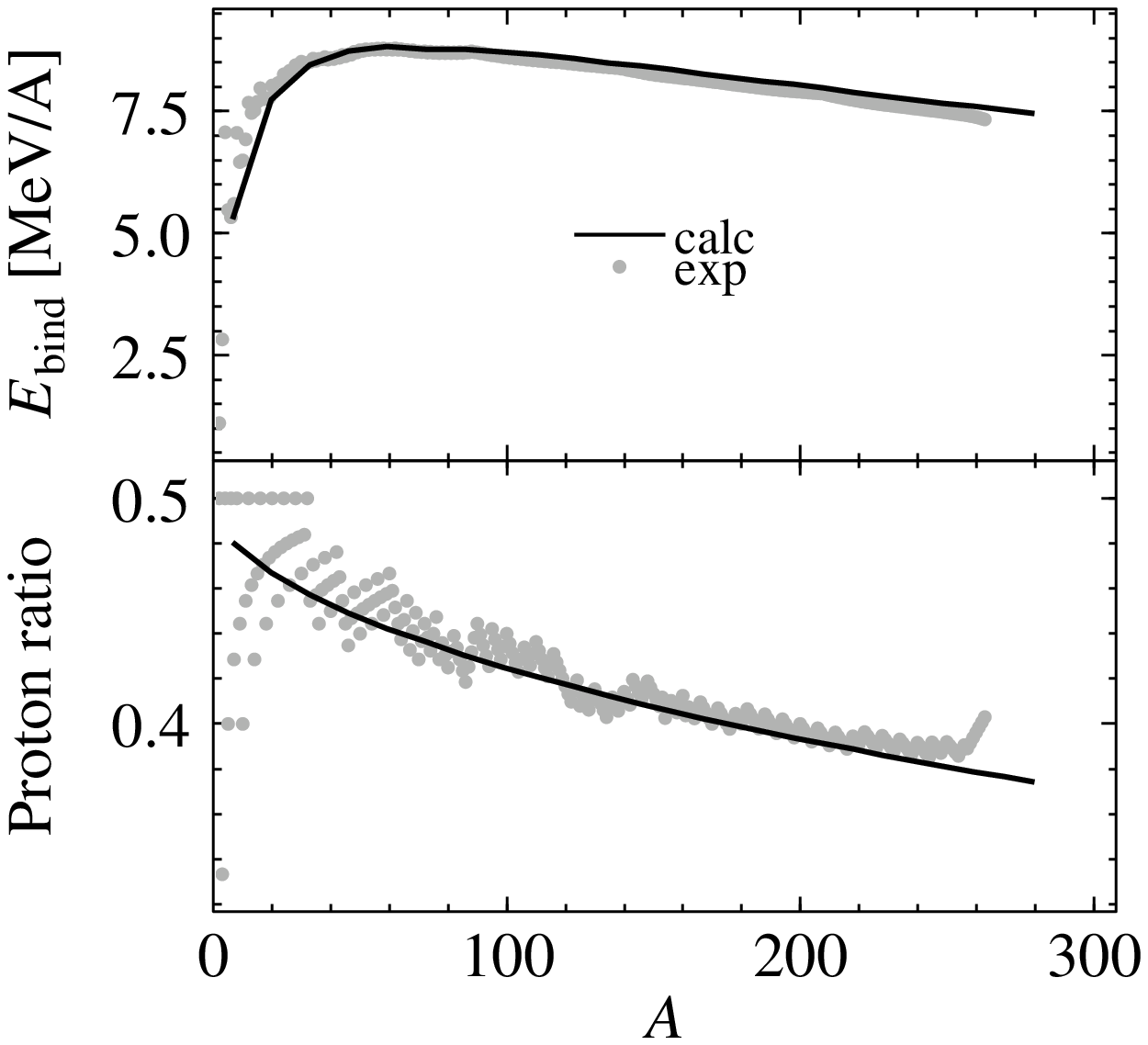}
  \caption{Properties of finite nuclei obtained with the present RMF model.
  Left: the density profiles of typical nuclei.
The proton number densities (solid curves) are compared with the experiment.
Right: binding energy per nucleon and the proton number ratio
of finite nuclei.
}
\label{finite}
\end{figure}

In Fig.~\ref{finite} (left panel) we show the density profiles of some typical nuclei.
One can see how well our framework may reproduce the density
profiles.
To get a still better fit, especially around the surface region,
we might need to include the derivative terms of the nucleon densities,
as we have already remarked.
Fine structures seen in the empirical density profiles,
which may come from the shell effects (see, e.g., a proton density  dip
at the center of a light $^{16}$O nucleus),
cannot be reproduced by the mean-field theory.
The effect is seen of
the rearrangement of the proton density distribution in heavy nuclei. Protons repel each
other, which enhances their contamination near the surface of heavy nuclei. This
effect is analogous to the charge screening effect for the Coulomb
potential in a sense that the proton
distribution is now changed not on the scale of  the nuclear radius, but
on another length scale, that we  will call the proton Debye screening
length, see Eq.\ (\ref{Deb}) below.
It gives rise to important consequences for the pasta structures since typically the
proton Debye screening 
length is less than the droplet size.
The optimal value of the proton ($Z$) to the total baryon ($A$)
number ratio $Y_p =Z/A$ is obtained by imposing the beta
equilibrium condition for a given baryon number. 
 Figure \ref{finite}
(right panel) shows the baryon number dependencies of the binding
energy per baryon and the proton number ratio. We can see that the bulk
properties of finite nuclei (density, binding energy, and proton
to baryon number ratio) are satisfactorily reproduced for our present
purpose.

Note that in our framework we must use a
sigma mass $m_{\sigma}=400$ MeV \cite{centelles93}, a slightly smaller value than
that one usually  uses, to get an appropriate fit. If we used a
popular value $m_\sigma\approx 500$ MeV, finite nuclei would be
over-bound by about 3 MeV/$A$. The actual value of the sigma
mass (as well as the omega and rho masses) has little relevance
for the case of infinite nucleon matter, since it enters the
thermodynamic potential only in the combination
$\widetilde{C}_{\sigma}=g_{\sigma N}/m_{\sigma}$. However meson
masses  are important characteristics of finite nuclei and of
other non-uniform nucleon systems, like those in pasta. The
effective meson mass characterizes the typical scale for the
spatial change of the meson field and consequently it affects the
value of the effective surface tension \cite{voskresurf}.

\section{Non-uniform structures in nucleon matter}\label{Non-uniform}

\subsection{Nucleon  matter at fixed proton number ratios}\label{Ypfix}

\begin{figure*}
\includegraphics[height=.39\textheight]{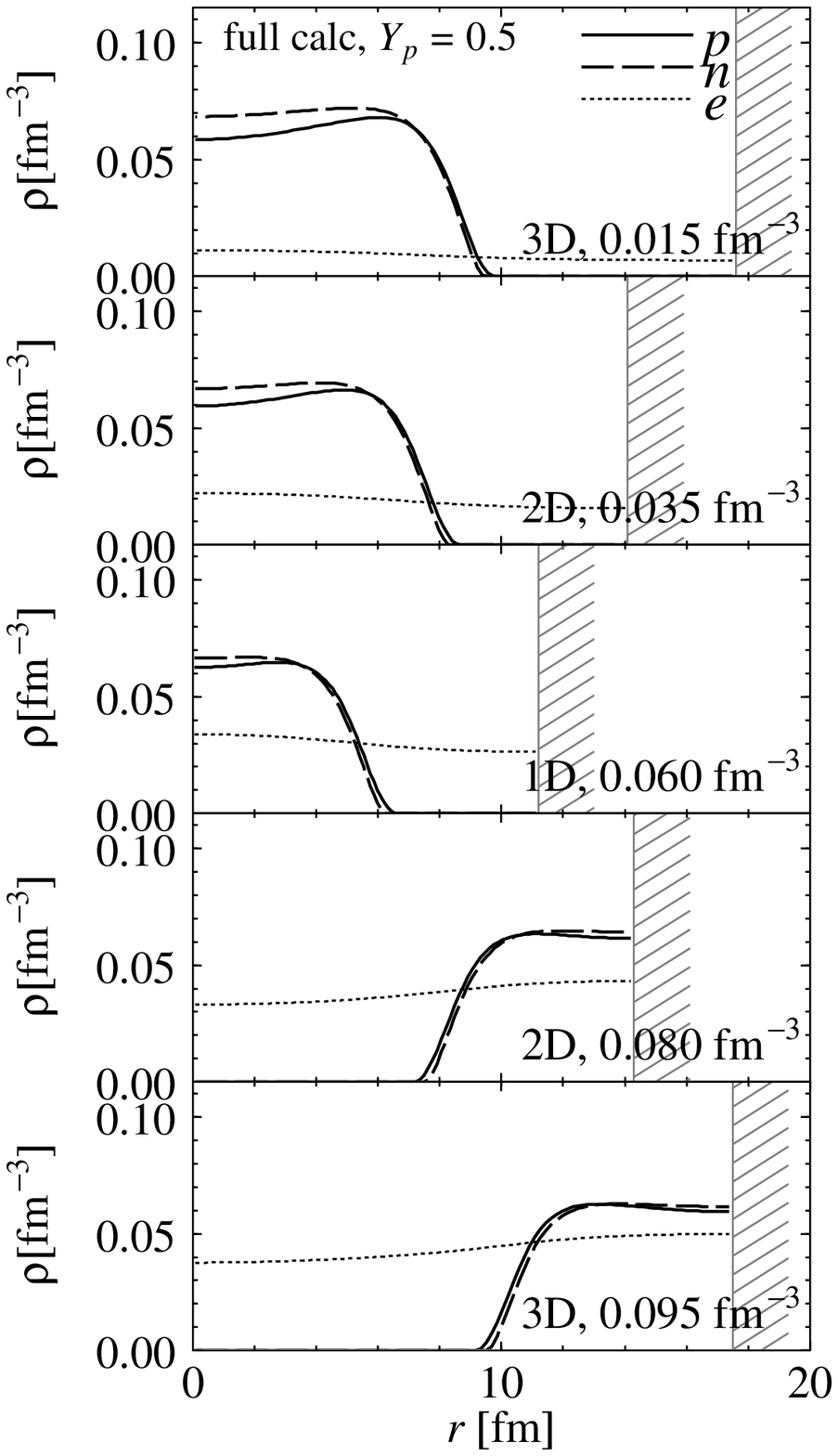}
\includegraphics[height=.39\textheight]{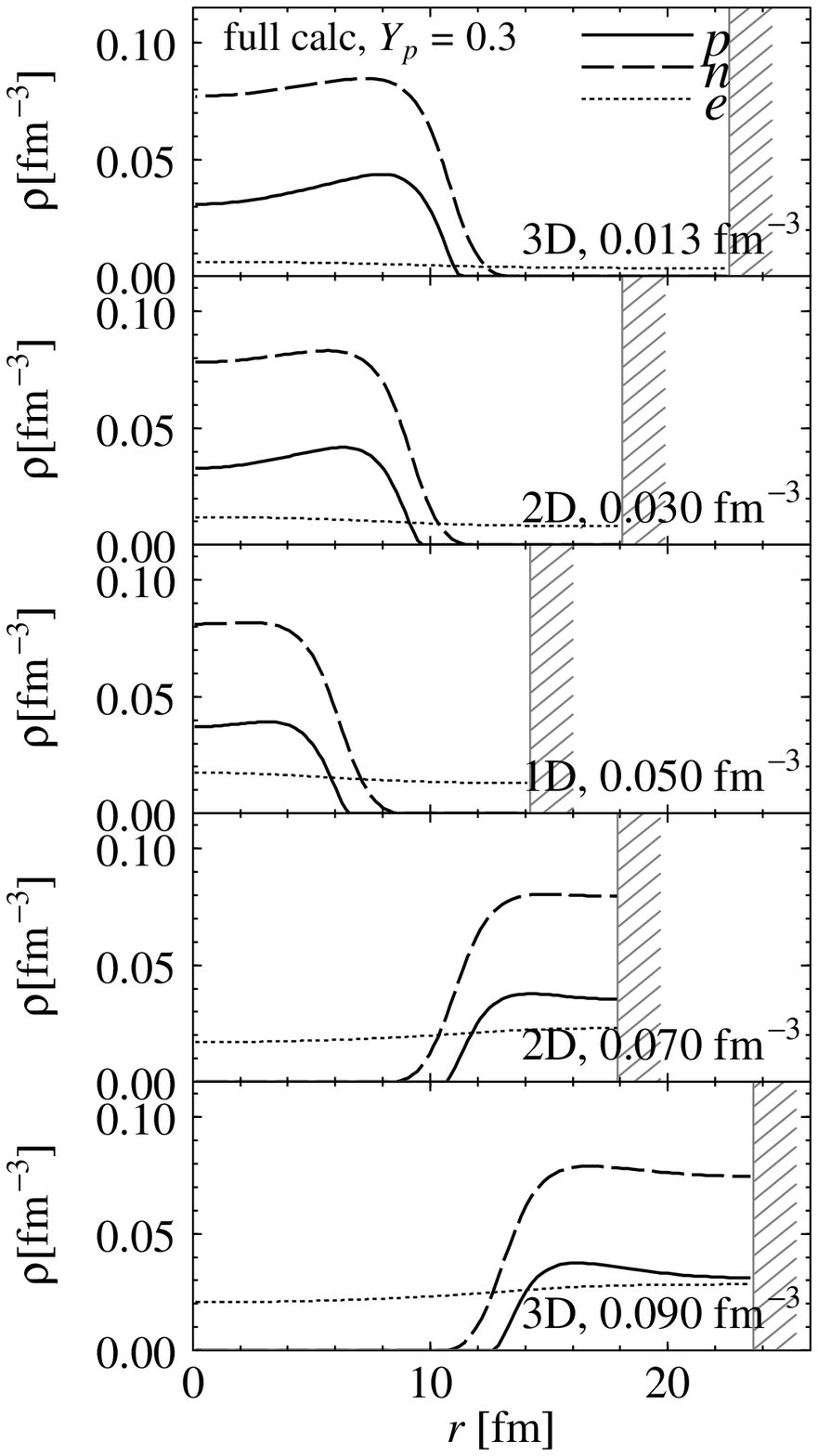}
\includegraphics[height=.39\textheight]{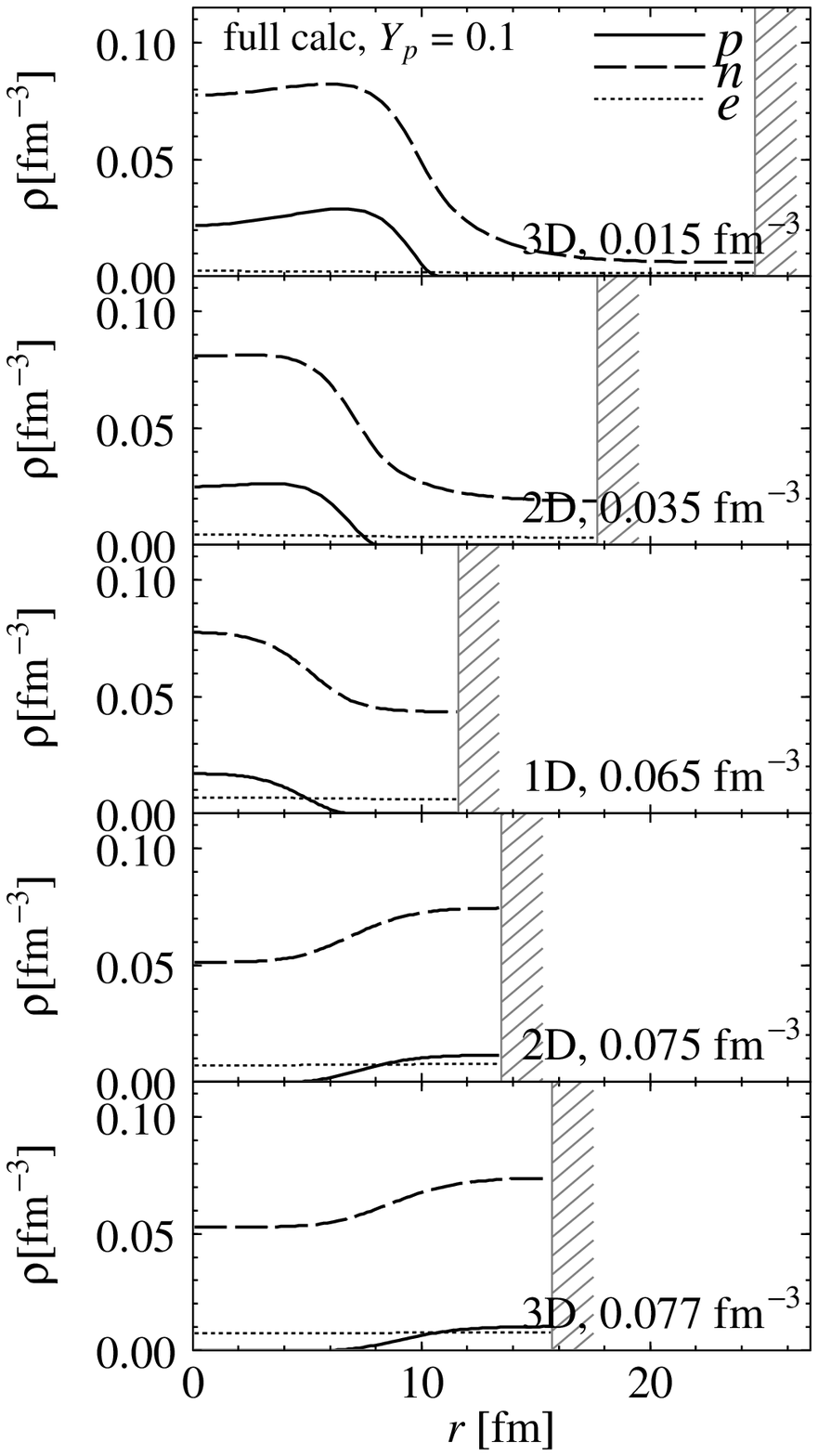}
\caption{
Examples of the density profiles in the cell for symmetric nuclear
 matter with $Y_p$=0.5 (left panel) and for asymmetric matter
 with $Y_p=0.3$ (center panel) and 0.1 (right panel). 
}
\label{proffixfull}
\end{figure*}
\begin{figure*}[t]
\includegraphics[width=.28\textwidth]{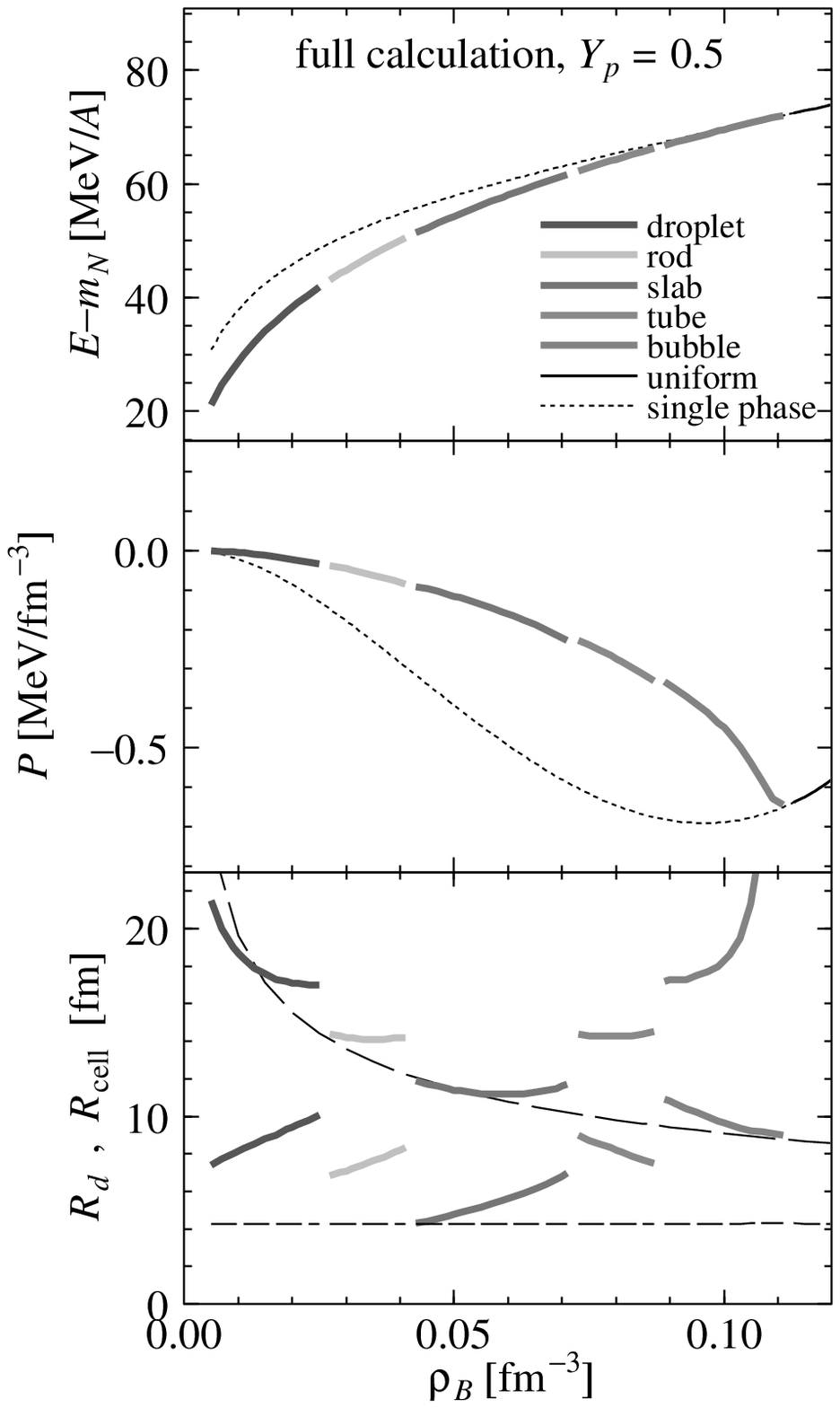}
\includegraphics[width=.28\textwidth]{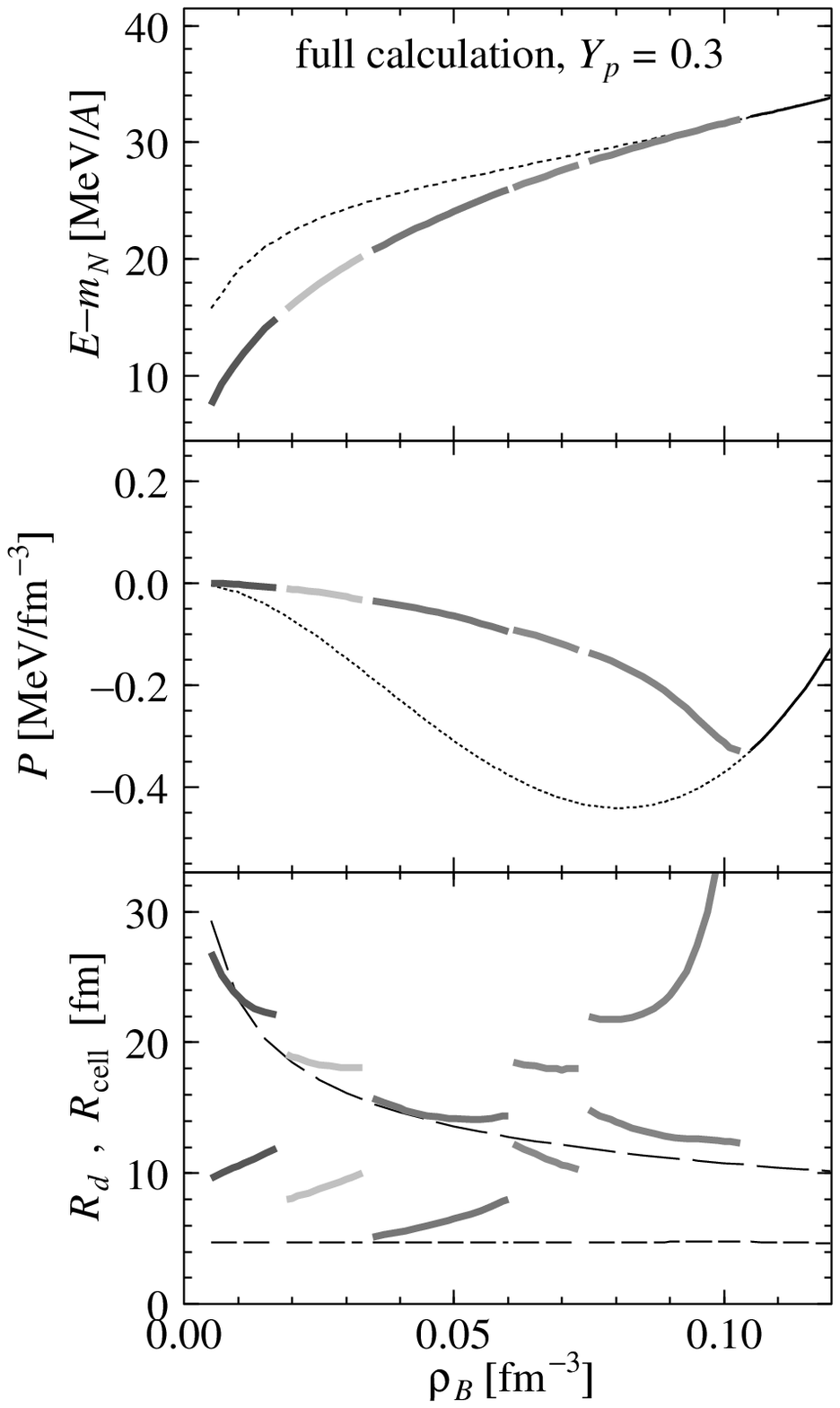}
\includegraphics[width=.28\textwidth]{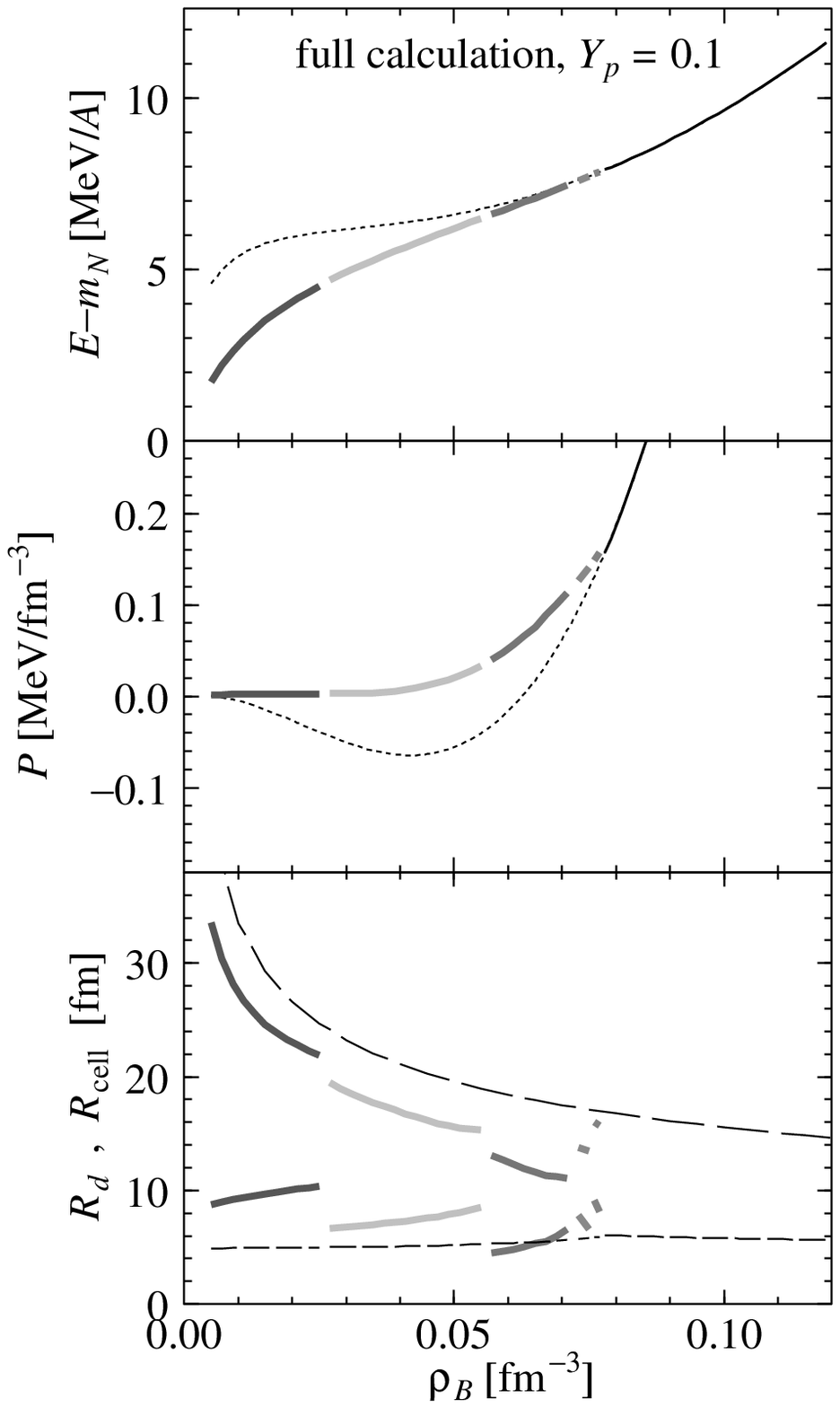}
\caption{
Binding energy per nucleon, baryon partial pressure, and the cell and nuclear sizes for symmetric nuclear
 matter with $Y_p$=0.5 (left panel), and for asymmetric matter with
 $Y_p=0.3$ (center panel) and 0.1 (right panel). 
}
\label{eosfixfull}
\end{figure*}

First, we are concentrated  on the discussion of the behavior of the nucleon
matter at a fixed value of the proton number ratio $Y_p$.
Particularly, we explore the proton number ratios $Y_p=0.1$, 0.3, and 0.5.
The cases $Y_p=0.3$ -- 0.5 should be relevant for the supernova matter
and for newly born neutron stars. Figure \ref{proffixfull} shows
some typical density profiles inside the Wigner-Seitz cells. The
geometrical dimension of the cell  is denoted as ``3D''
(three-dimensional sphere), etc. The horizontal axis in each panel
denotes the radial distance from the center of the cell. The cell
boundary is indicated by the hatch.
 From the top to the bottom the configuration corresponds to
droplet (3D), rod (2D), slab (1D), tube (2D), and bubble (3D).
The nuclear ``pasta'' structures are clearly manifested.
For the lowest $Y_p$ case ($Y_p =0.1$), the neutron density is finite at any point:
the space is filled by dripped neutrons;
the neutron-drip value of $Y_p$ 
is, e.g.,  around 0.26 in our 3D calculation.
For a higher  $Y_p$, the neutron density
drops to zero outside the nucleus. The proton number density always drops to zero outside the nucleus.
We can see that the charge screening effects are pronounced. Due to the spatial rearrangement of electrons the
electron
density profile becomes no more uniform. This non-uniformity of the electron distribution is
more pronounced for a higher $Y_p$ and a higher density. Protons repel each
other.
Thereby the proton density profile substantially deviates from the step-function. The proton
number is enhanced near the surface of the nucleus.

The equation of state (EOS) for the sequence of geometric structures is 
shown in Fig.~\ref{eosfixfull} (top panels) 
as a function of the averaged baryon-number density.
Note that the energy $E-m_N$ also includes the
kinetic energy of electrons, which makes the total pressure
positive. The lowest-energy configurations are selected among
various geometrical structures. The most favorable configuration
changes from the droplet to rod, slab, tube, bubble, and to the
uniform one (the dotted thin curve) with an increase of density.
The appearance of non-uniform structures in matter results in a
softening of EOS: the energy per baryon
gets lower up to about 15 MeV$/A$ compared to the uniform matter.

 The middle panels in Fig.~\ref{eosfixfull} are partial
pressures without electron contribution versus averaged 
baryon number density. If electron partial pressure is included, the
total pressure becomes positive at all densities.

The bottom panels in Fig.~\ref{eosfixfull} are the cell radii 
$R_{\rm cell}$ and nuclear radii $R_{d}$ versus averaged 
baryon number density. The radius $R_{d}$ is defined by way of a
density fluctuation as
\begin{eqnarray}
R_{d}&=&
\cases{
\displaystyle
R_{\rm cell} { \langle\rho_p\rangle^2 \over \langle\rho_p^2\rangle },\ \ \ \hbox{(for droplet, rod, and slab)}\cr
\displaystyle
R_{\rm cell} \left(1-{\langle\rho_p\rangle^2 \over \langle\rho_p^2\rangle}\right),\ \ \ \hbox{(for tube and bubble)} }
\end{eqnarray}
where the bracket ``$\langle \rangle$'' indicates the average along
the radial (for 3D and 2D cases) or perpendicular (1D) direction in the cell.
Dashed curves show the Debye screening lengths of
electron and proton calculated as
\begin{eqnarray}\label{Deb}
\lambda^{(e)}_D&=&\left(-4\pi e^2{d\rho_e^{\rm av}\over d\mu_e}\right)^{-1/2},\\
\lambda^{(p)}_D&=&\left(4\pi e^2{d\rho_p^{\rm av}\over d\mu_p}\right)^{-1/2},
\end{eqnarray}
where $\rho_p^{\rm av}$ is the proton number density
averaged inside the nucleus (the region with finite $\rho_p$)
and $\rho_e^{\rm av}$ is the electron charge density averaged inside the cell.
Actually doing more carefully we should introduce four Debye screening lengths
$\lambda^{(i,<)}_D$ and $\lambda^{(i,>)}_D$
with a separate averaging for the interior and the exterior of the nuclei.
However we observe that the proton number density is always zero
in the exterior region and $\lambda^{(p,>)}_D =\infty$ thereby.
For electrons $\lambda^{(e,<)}_D$
and $\lambda^{(e,>)}_D$ are in general different
but both being large and of the same order of magnitude in
the pasta case under consideration. Therefore we actually
do not need a more detailed analysis of these quantities.
Note that these values are obviously gauge invariant.
Numerically, the cell radii $R_{\rm cell}$ for droplet, rod, and slab
configurations at $Y_p=0.5$ and 0.3 were proven to be close to
the electron screening length. For the tube, $R_{\rm cell}$ is
larger than ${\lambda}^{(e)}_D$. For $Y_p=0.1$, in all cases  $R_{\rm cell}$ is substantially
smaller than ${\lambda}^{(e)}_D$ and  the electron screening should be much weaker, thereby.
In all cases, except for bubbles (at $Y_p=0.5$ and 0.3), the structure radii $R_{d}$ are
smaller than ${\lambda}_D^{(e)}$.
This means that the Debye screening effect of electrons inside these structures should not be
pronounced. For bubbles at $Y_p=0.5$ and 0.3, ${\lambda}_D^{(e)}$
is substantially smaller than the cell size and the electron screening should
be significant, see Fig.\ \ref{phdiagcompare} below.
For $Y_p=0.5$, 0.3, 0.1 in all  cases (with the only exception $Y_p=0.1$ for
slabs), the value ${\lambda}_D^{(p,<)}$ is shorter than $R_{d}$.
Hence the density rearrangement of protons is essential
for the pasta structures, as it is indeed seen from the Fig.\ \ref{proffixfull}.

Knowing the baryon number density and the nuclear radius from
Fig.~\ref{eosfixfull}, one may estimate the atomic number of the
nucleus. In the case of droplets and for $Y_p=0.5$ the atomic
number of the droplet is $\simeq 25$ in the low density
limit and $\simeq 65$ at the maximum density of the droplet phase
$\rho_{B,d}^{(\rm max)} \simeq 0.025\ {\rm fm}^{-3}$.

\begin{figure*}
\begin{minipage}{0.48\textwidth}
\includegraphics[height=.39\textheight]{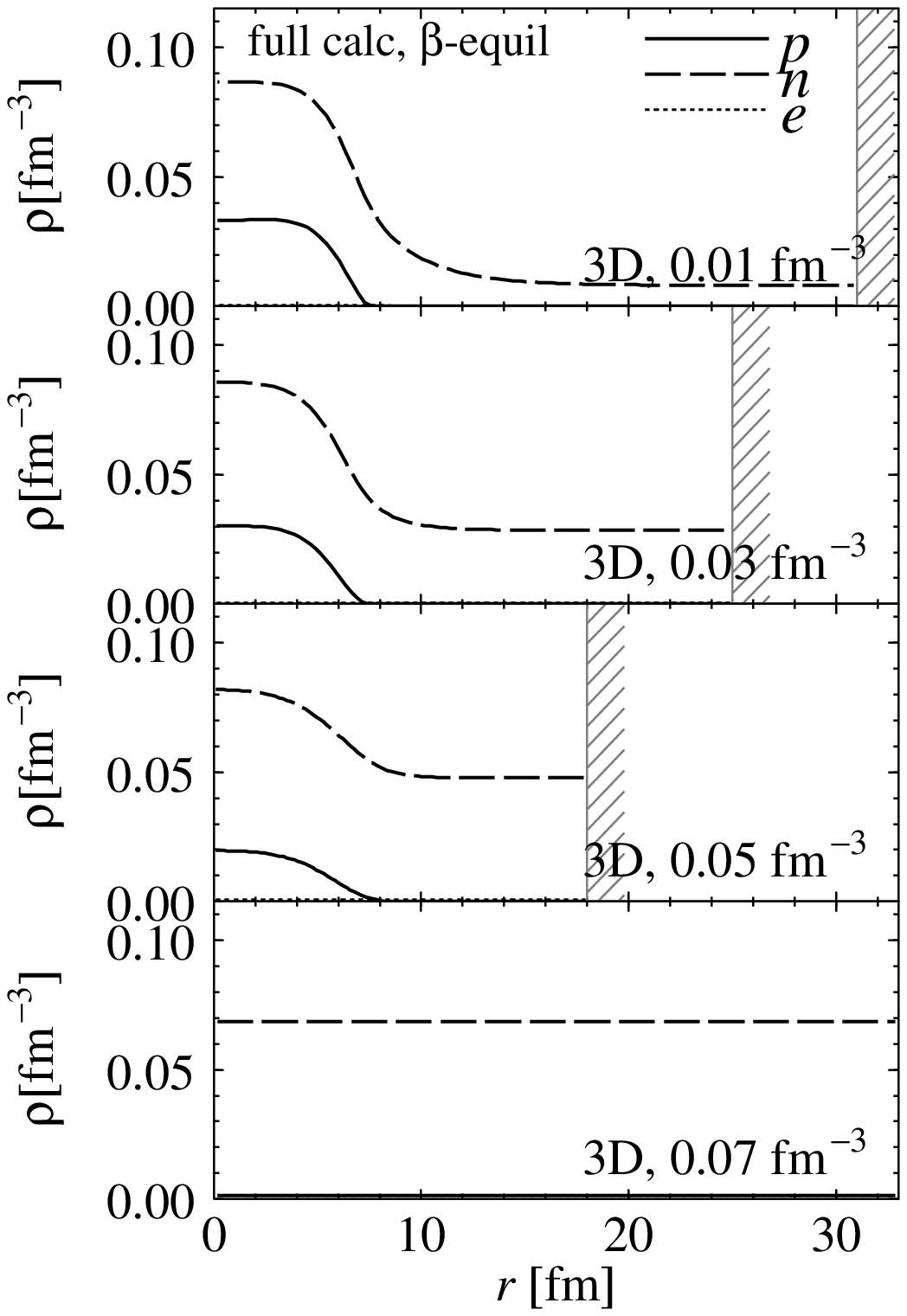}
\caption{
Density profiles in the cell for nuclear matter in beta equilibrium with
 baryon-number densities, 0.01, 0.03, 0.05 and 0.07 fm$^{-3}$ from top to bottom.
}
\label{profbeta}
\end{minipage}
\hspace{\fill}
\begin{minipage}{0.48\textwidth}
\includegraphics[width=.7\textwidth]{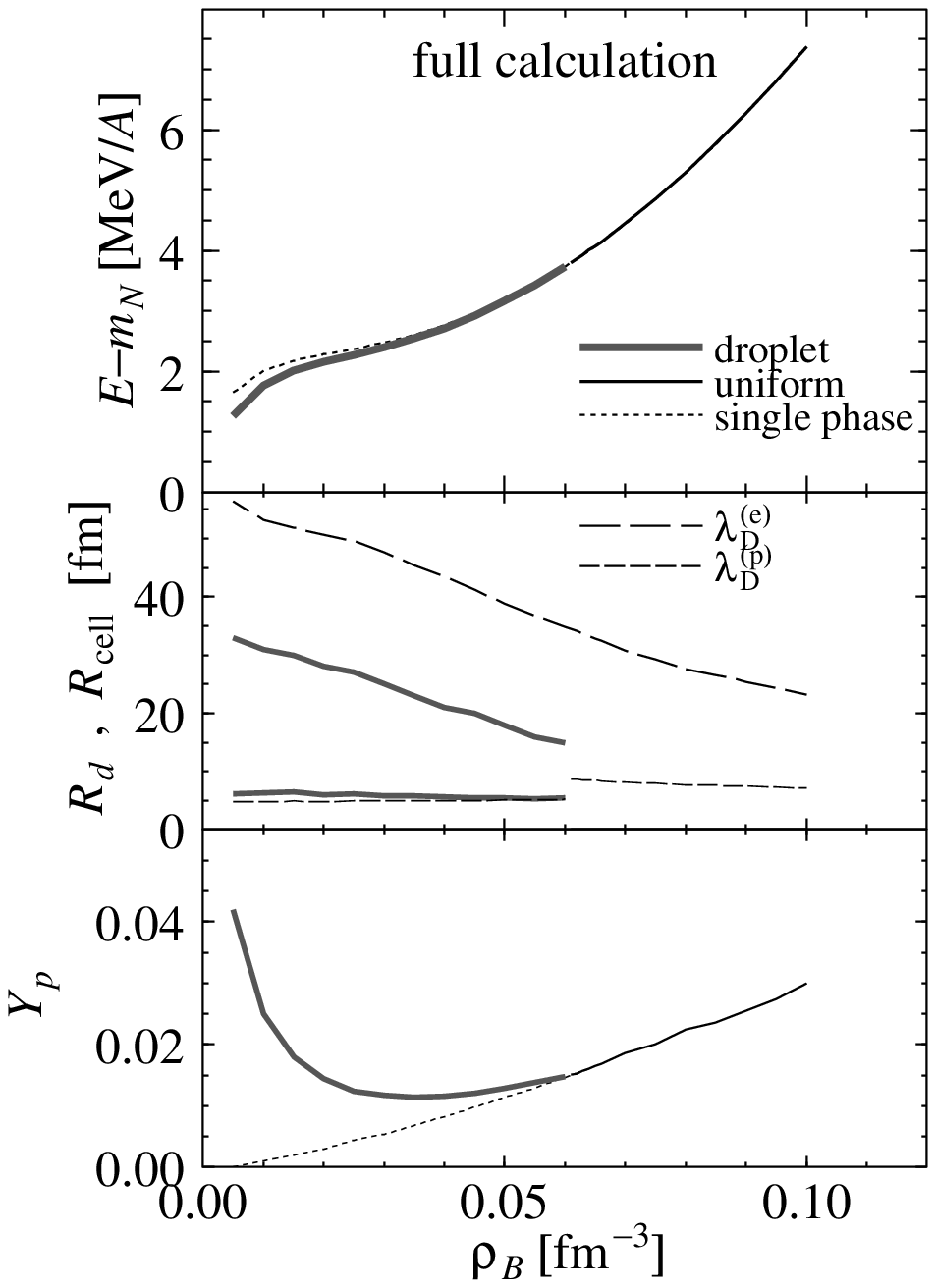}
\caption{
Binding energy (top), cell and nuclear sizes (middle), and proton number ratio (bottom)
in the cell for nuclear matter in beta equilibrium.
}
\label{eosbeta}
\end{minipage}
\end{figure*}

\subsection{Nucleon matter in beta equilibrium}\label{BetaEq}

Next, we  consider the neutron star matter at zero temperature, and 
explore the non-uniform structures for the nucleon matter in beta
equilibrium. Figure \ref{profbeta} shows the density profiles for
different baryon number densities. The droplet structure itself is
quite similar to the case of the fixed proton number 
ratio $Y_p=0.1$ considered above. The apparently different feature in this case is
that only the droplet configuration appears as a non-uniform
structure. It should be noticed, however, that the presence or 
absence of the concrete pasta structure sensitively depends on the
choice of the effective interaction. 

In Fig.~\ref{eosbeta} we plot
the energy per baryon (top), the cell and nuclear sizes (middle), and the proton
number ratio (bottom).
The effect of the non-uniform structure on EOS (the difference between
the energy of uniform matter and that of non-uniform one) is small.
However, the proton number ratio is significantly affected by the presence of the pasta at lower densities.
In the zero-density limit, the proton number ratio should converge to that of
the normal nuclei.
The droplet radius and the cell radius in the middle panel of Fig.~\ref{eosbeta}
are always smaller than the electron Debye screening length $\lambda_D^{(e)}$.
Thereby the effect of the electron charge screening is small.
Unlike the fixed $Y_p$ cases, 
the droplet radius is comparable to the proton Debye screening length, which means that 
the effect of the proton rearrangement is not pronounced in this case.
In fact, there is no enhancement of the proton number density near the surface in Fig.\ \ref{profbeta},  
in contrast to Fig.\ \ref{proffixfull}.

\section{Comparison with other calculations}\label{CompCalc}

\begin{figure*}
\includegraphics[width=.28\textwidth]{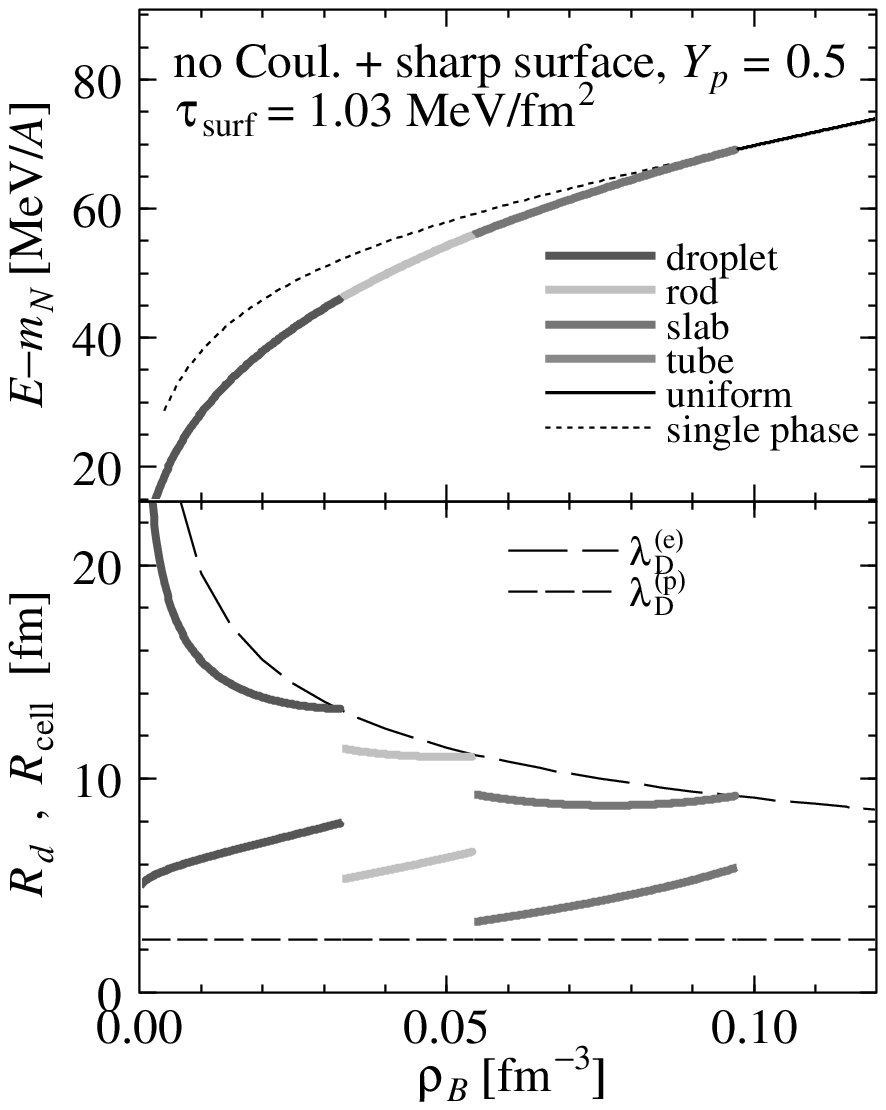}
\includegraphics[width=.28\textwidth]{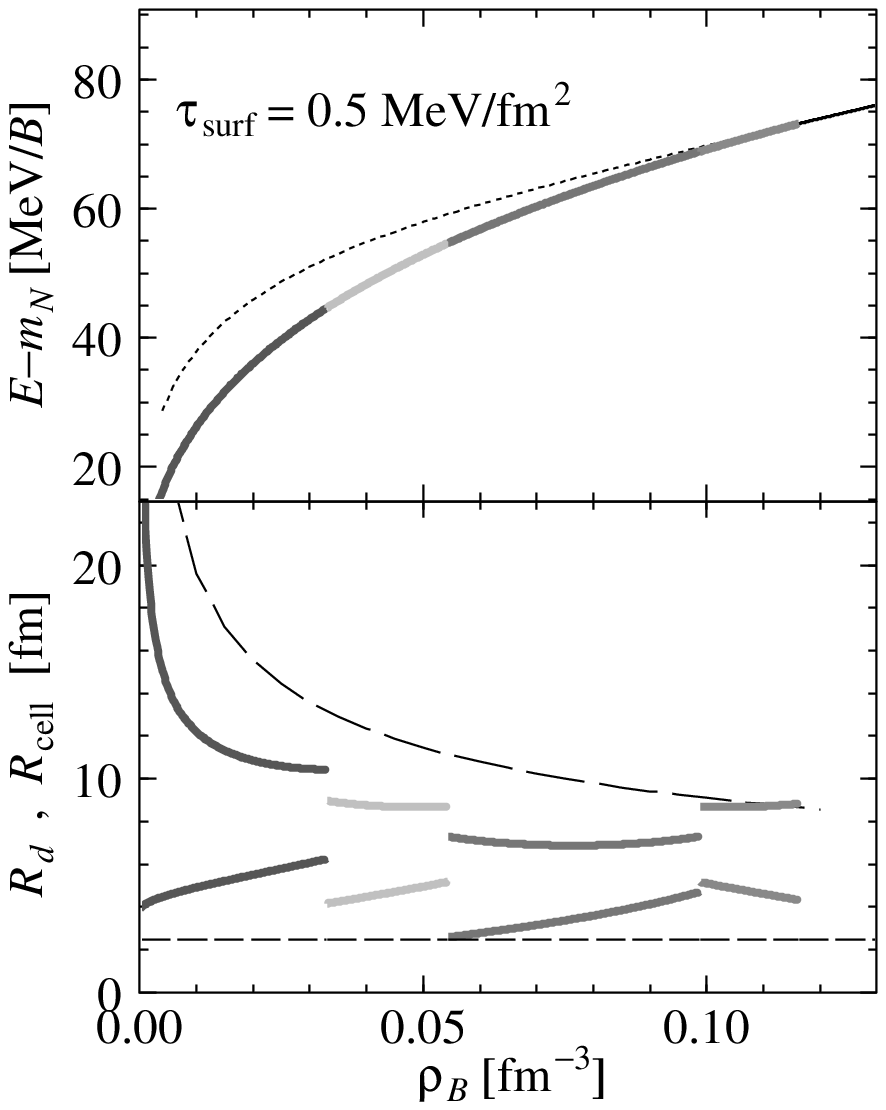}
\caption{ Bulk calculations with the surface tension parameter
$\tau_{\rm surf}=1.03$ and 0.5 MeV/fm$^2$.
} \label{figBulk}
\end{figure*}

\begin{figure*}
\includegraphics[height=.39\textheight]{Prof-05BW.ps}
\includegraphics[height=.39\textheight]{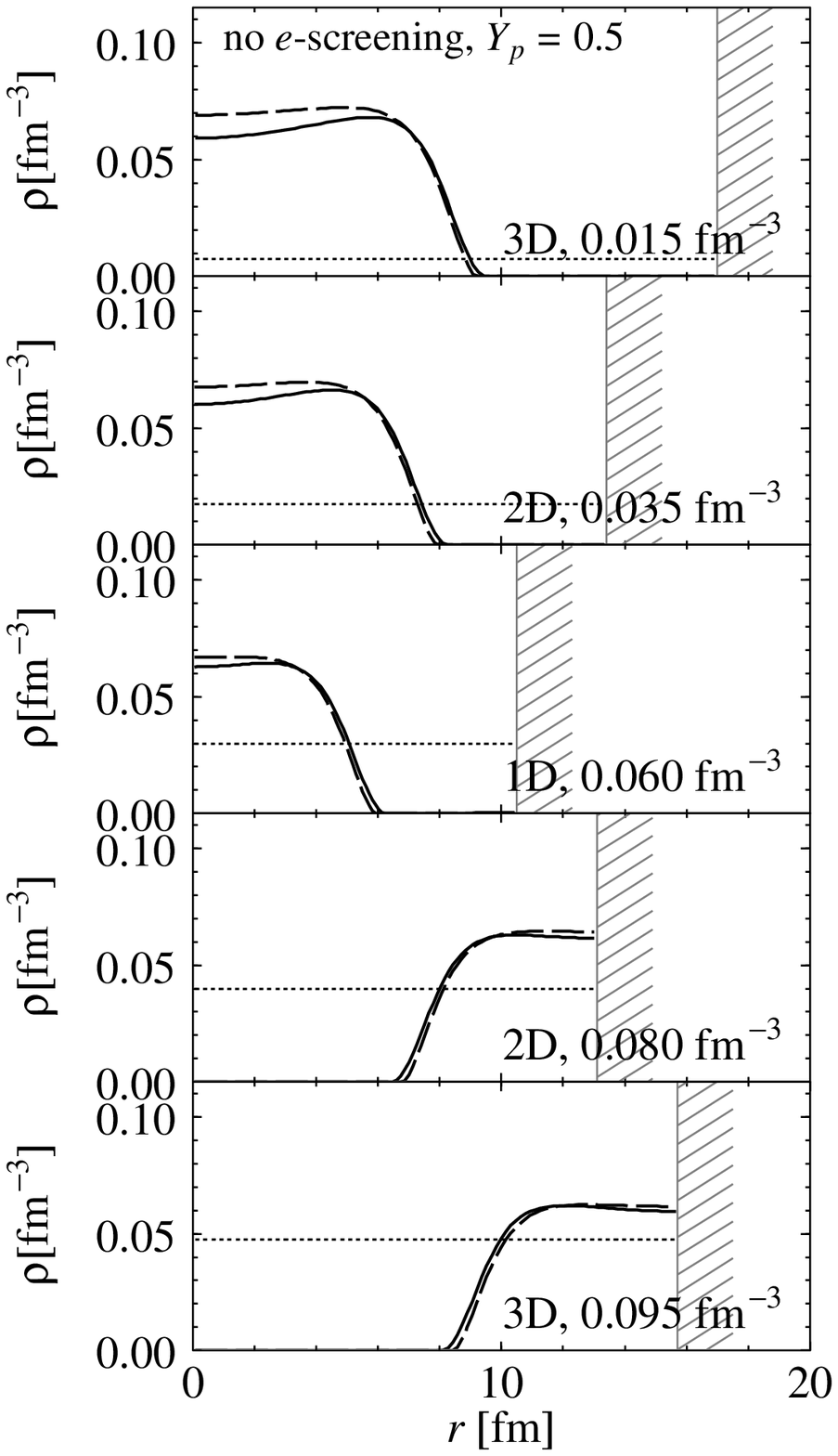}
\includegraphics[height=.39\textheight]{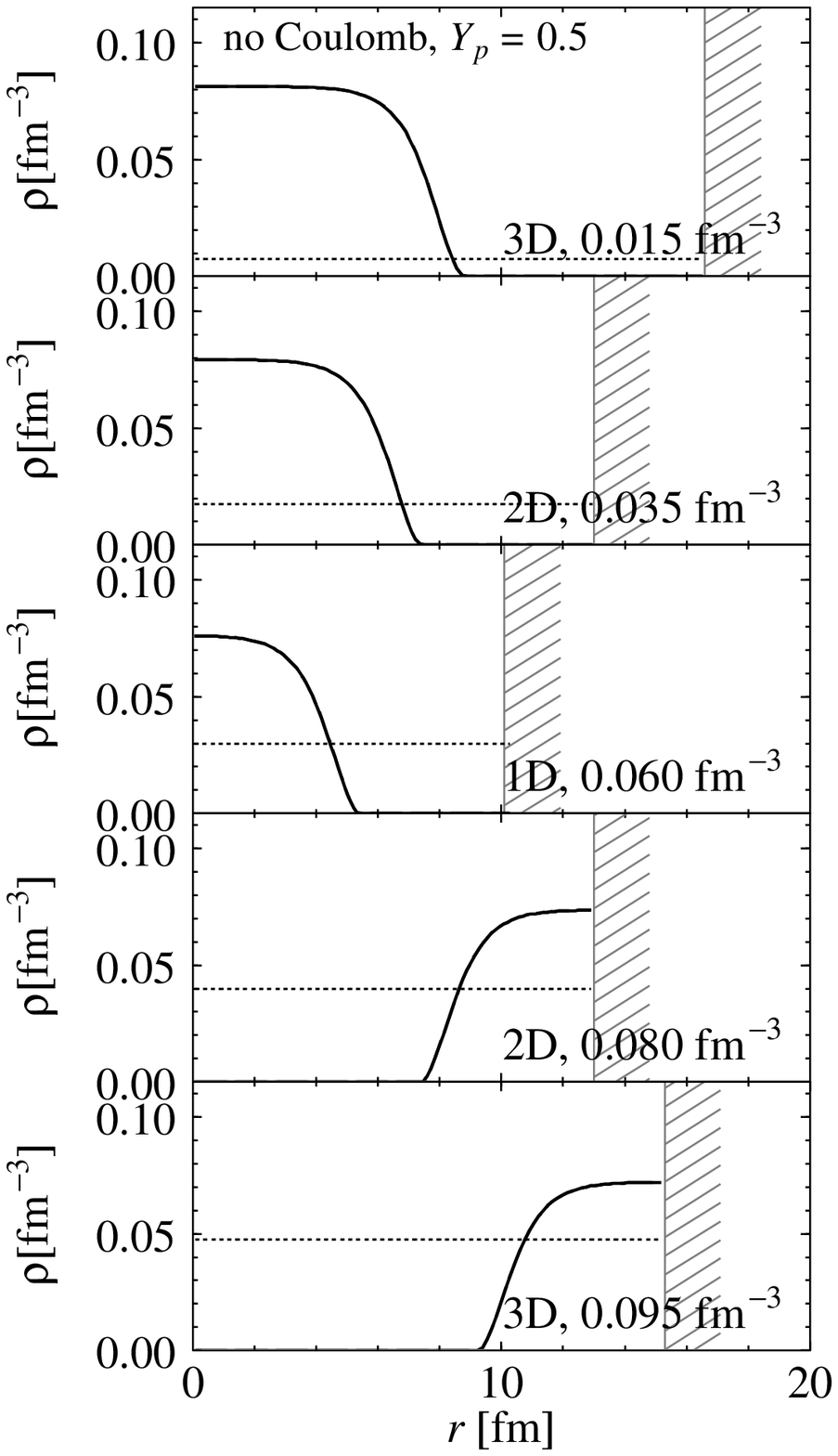}
\caption{
Comparison of the density profiles for
different treatments of the Coulomb interaction.
 From the left: full calculation, without electron screening,
and ``no Coulomb'' calculation. The proton number ratio is $Y_p$=0.5 for all cases. 
}\label{profcompare}
\end{figure*}

\begin{figure*}
\includegraphics[width=.28\textwidth]{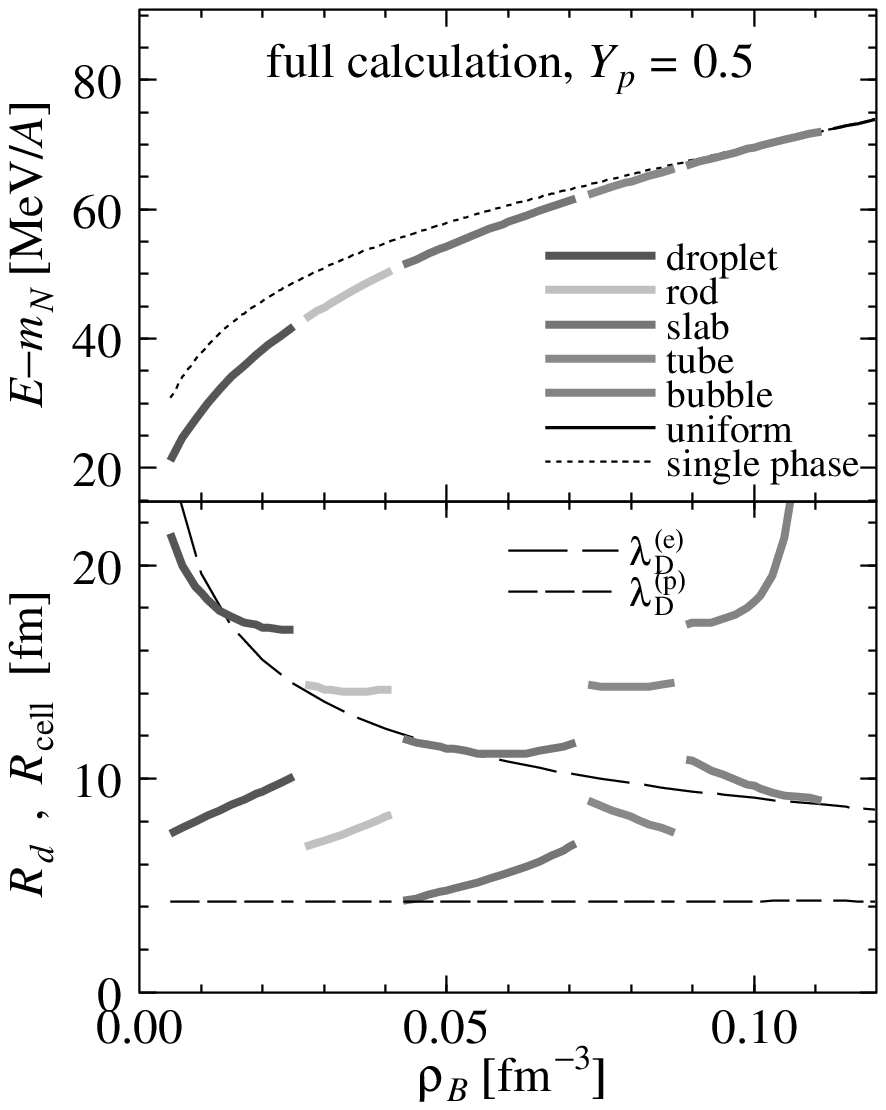}
\includegraphics[width=.28\textwidth]{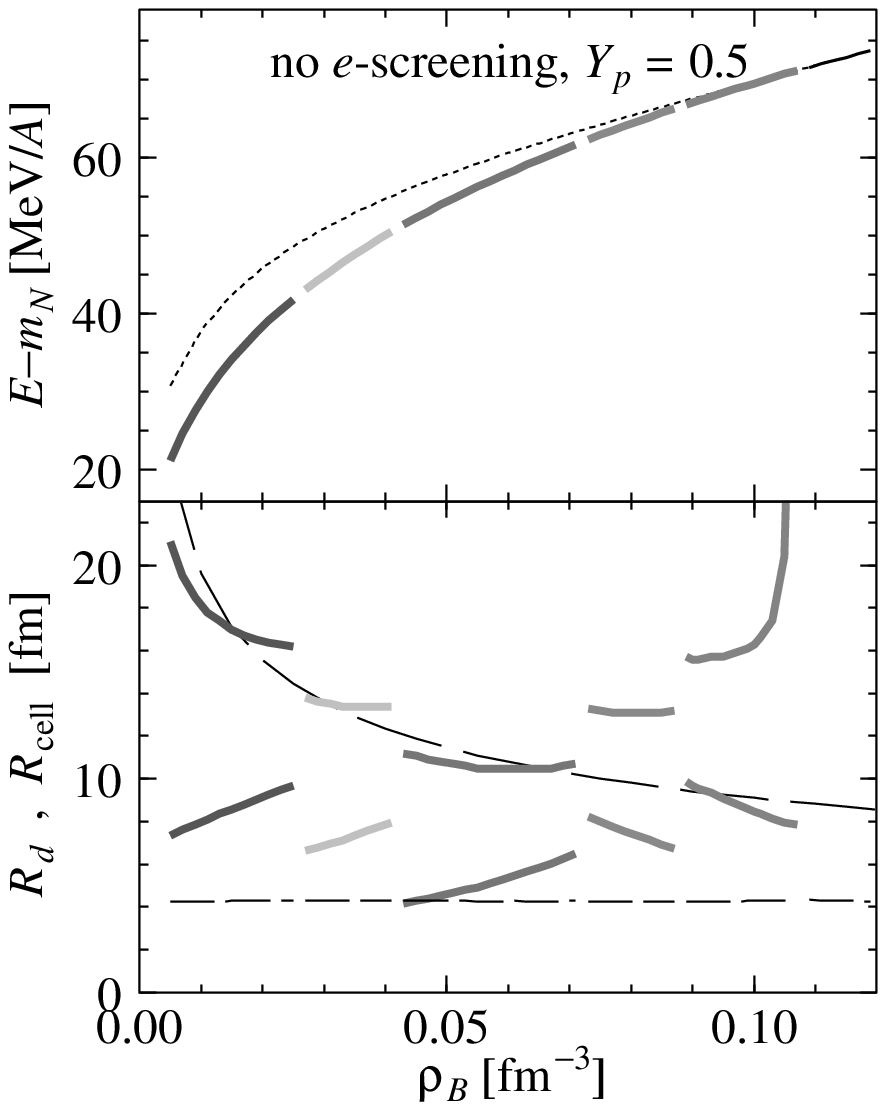}
\includegraphics[width=.28\textwidth]{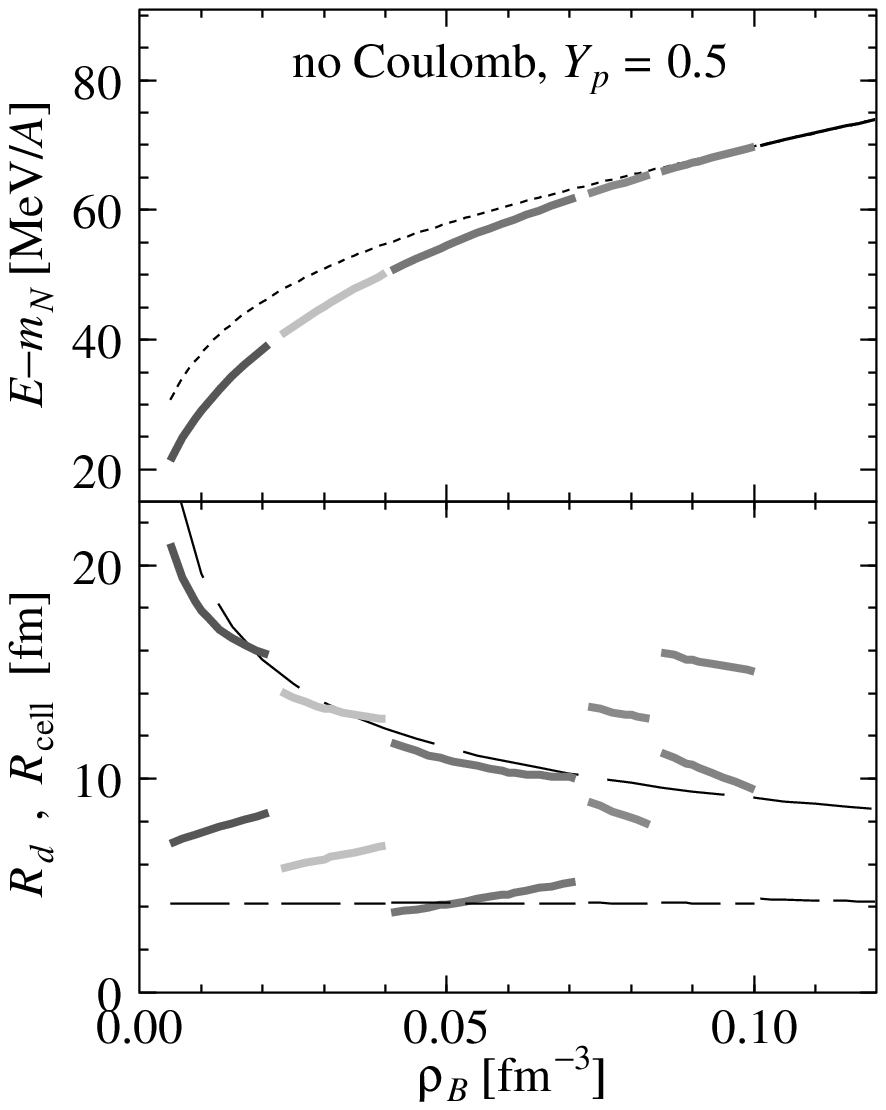}
\caption{
Comparison of the density profiles for
different treatments of the Coulomb interaction.
 From the left: ``full'' calculation, ``no electron screening'',
and ``no Coulomb'' calculation.
The proton number ratio is $Y_p$=0.5 for all cases.
}
\label{eoscompare}
\end{figure*}

\begin{figure}
\includegraphics[width=.48\textwidth]{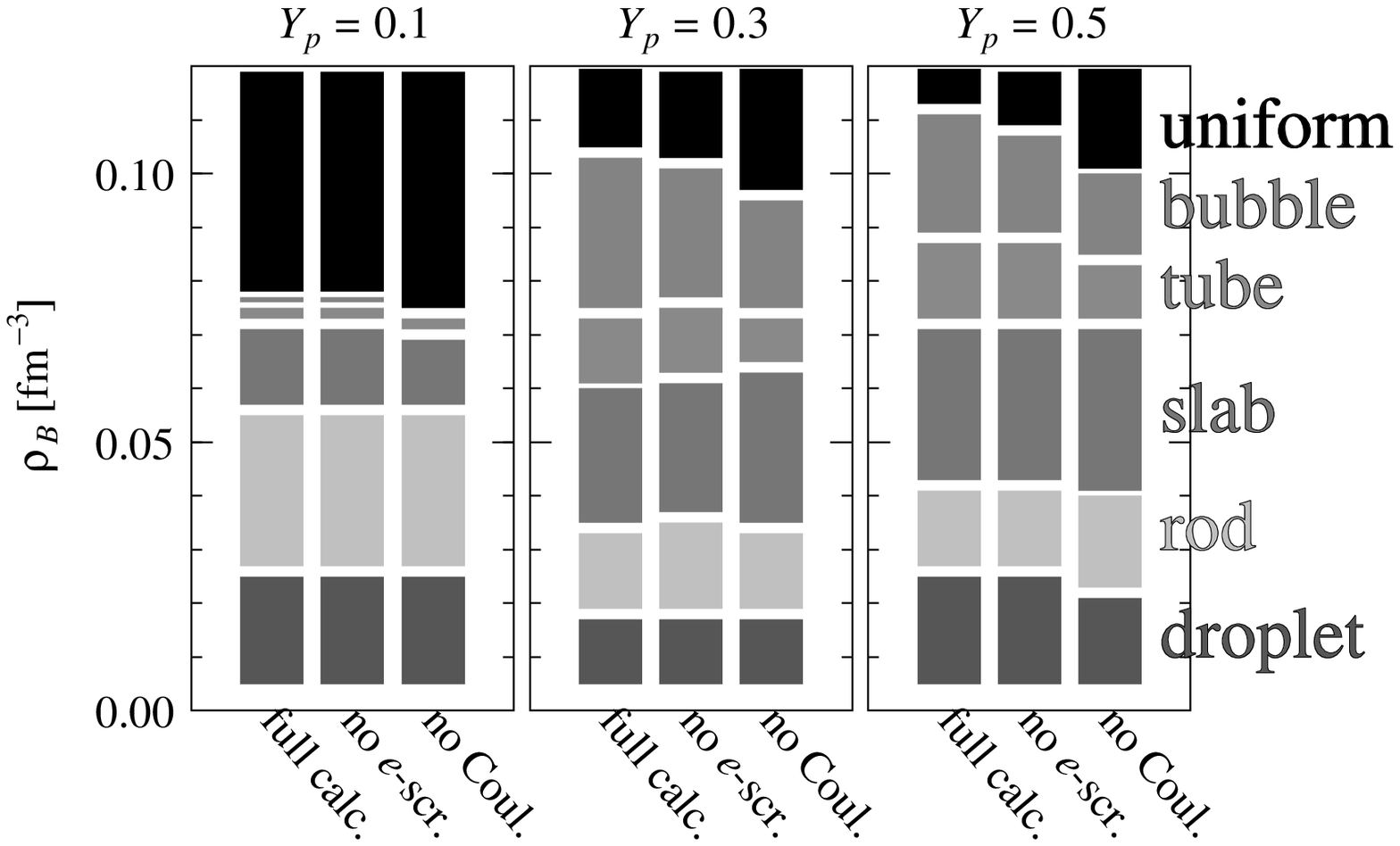}
\caption{
Comparison of the phase diagrams between
different treatments of the Coulomb interaction.
}
\label{phdiagcompare}
\end{figure}

 In this section we compare our DFT calculation with
others to explore the effects of the surface, the charge
rearrangement, and the fully consistent treatment of the 
density distribution.

First let us focus on  a very simplified treatment that has
been used in the literature. We consider  the bulk calculation
supplemented by a simplified  treatment of the finite-size effect.
For description of the latter we introduce   the surface tension
and a bare (non screened) Coulomb interaction. This calculation
assumes a sharp boundary between dense and dilute phases,
uniform baryon density distribution inside each phase, and uniform
electron density distribution all over the cell. To further
specify this approximation we use the term ``no Coulomb + sharp
surface''. We totally discard the Coulomb potential in equations of
motion and drop the Poisson equation (``no Coulomb'') and we reduce
the mean fields to their constant bulk values in the interior and
the exterior of the structure (``+ sharp surface''). The Coulomb
energy, being evaluated with the step-function-like  density
profiles, and the surface energy, being expressed via the surface
tension parameter $\tau_{\rm surf}$ are added to the total bulk
energy.

The volume fraction of each phase is simply calculated without
taking into account of the finite-size effect (``bulk
calculation'').
Details of the ``no Coulomb + sharp surface'' calculation
are presented in Appendix \ref{app:bulk}.

Figure \ref{figBulk} shows the EOS obtained by the ``no
Coulomb + sharp surface'' calculation performed at different
values of the surface tension.
In this case for $Y_p=0.5$, the dilute phase includes no
baryon. The value of the surface tension parameter
$\tau_{\rm surf}\simeq 1.03$~MeV$/$fm$^{2}$ fits the liquid-drop
binding energies of finite nuclei.
Note that the appearance or disappearance of the pasta
structure  essentially depends on the value of the surface
tension. With a larger  value of the surface tension, the density
region of the pasta structure reduces and even some of the
structures, e.g. ``tube'' and ``bubble'', disappear. With a
smaller value of the surface tension the region of the pasta
structure broadens and all kinds of pasta structures appear if we take
$\tau_{\rm surf}\leq 0.3$ MeV/fm$^2$. However, if we put surface
tension zero, the mixed phase reaches up  from zero to the
saturation density $\rho_0$ without any specific geometry.
Therefore, from the given  example  we see that the surface
tension plays a crucial role in the appearance of pasta
structures.
Remember that, in the case under consideration, the pasta
structures are realized by a balance of the surface tension and
the bare Coulomb interaction, which reads $E_{\rm surf} =2E_{\rm
Coul}$, where $E_{\rm surf}$ is the surface energy and $E_{\rm
Coul}$ is the bare Coulomb energy. Therefore, the Coulomb
interaction is important, as well. Please also note that the
surface tension introduced here simulates effects of the spatial
changes of the meson mean-fields. In our ``full calculation'' the
latter  effects are taken into account explicitly whereas  purely
``bulk calculations'' completely disregard these effects.

Next we compare three kinds of calculations with different
treatments of the Coulomb interaction. One is the ``full
calculation'' which we have done here. The second is the
calculation that disregards  electron screening (``no
$e$-screening''): a constraint is used that the electron density
should be uniform. In this calculation the Coulomb potential
$V_{\rm Coul}$ in Eq.~(\ref{eq:rhoe}) is replaced by a constant
$V_0 =0$,
\begin{equation}
{\rho_e}=-(\mu_e-V_{0})^3/3\pi^2. \label{Vnul}
\end{equation}
In the full calculation the value of $V_0$ is arbitrary, and one
can take $V_0$ for the sake of convenience, e.g.\ as
$V_0=0$, either set it equals the averaged value of
$V_{\rm Coul}(r)$ over the cell: recall that $V_{\rm Coul}$ either
$\mu_e$ alone do not have a physical meaning but only the
combination $\mu_e-V_{\rm Coul}$ is meaningful due to the gauge
invariance, cf.\ \cite{voskre,voskre1}. However in the case ``no
$e$-screening'' the gauge invariance is violated as it follows from
Eq.\ (\ref{eq:cpotBp}), since we replace  $V_{\rm Coul}$ to  $V_0
=0$ in the equation for the electron chemical potential but remain
$V_{\rm Coul}$ in the equation for the proton chemical potential
and thus in the expression for the proton number density. We do this
procedure just to demonstrate the efficiency of the proton
rearrangement artificially suppressing that of the electron one.
The third calculation ``no Coulomb'' is performed by totally
discarding the Coulomb potential $V_{\rm Coul}$ in equations of
motion. Accordingly, the Poisson equation is discarded as well.
After getting the density profiles, the Coulomb energy, being
evaluated using charge densities thus determined, is added to the
total energy.
This calculation is similar to the above discussed ``no
Coulomb + sharp surface'' calculation. The difference is that the
effect of the density variation  near the structure surface is
automatically incorporated in explicit ``no Coulomb'' calculation,
while in the above ``no Coulomb + sharp surface'' calculation this effect is 
hidden in the value of the surface tension.

In Fig.~\ref{profcompare} compared are
the density profiles for
different treatments of the Coulomb interaction.
The left panel is the same as
that in Fig.~\ref{proffixfull}. It demonstrates the ``full'' calculation.
It seems that there is almost no difference between
the nucleon density of the ``full'' calculation and that
of ``no $e$-screening'' calculation (center).
The case of ``no Coulomb'' calculation (right), contrarily,
shows a significant difference especially in the proton number density.
The reason is simple: the electron Debye screening length is large, whereas the
proton Debye screening length is rather short. Thus the proton screening effects are
much more pronounced than the electron ones.

The EOS as a whole (upper panels in Fig.~\ref{eoscompare}) shows
almost no dependence on the treatments of the Coulomb interaction.
This agrees with a general statement that the variational
functional is always less sensitive to  the choice of the trial
functions than the quantities linearly depending on them. 
Nevertheless, sizes of the cell and the nucleus (lower panels in
Fig.~\ref{eoscompare}) especially for tube and bubbles are
different. In the cases of the ``full calculation'' and ``no
$e$-screening'', the cell radii of ``tube'' and ``bubble''
structures and that of ``slab'' structure get larger with an
increase of density, while they are monotonically decreasing in
the case of ``no Coulomb'' calculation. We see almost no
differences between the ``full'' and ``no $e$-screening''
calculations that again demonstrates relevance of the proton
screening and weakness of the electron screening effects. The only
significant difference remains for bubbles as seen from
Fig.~\ref{phdiagcompare}. The other effect illustrated by
Fig.~\ref{phdiagcompare} is a  difference in the density range for
each pasta structure. The ``full'' treatment of the Coulomb
interaction slightly increases the region of the nuclear pasta.
For $Y_p =0.1$ the differences between ``full''  and ``no
$e$-screening'' calculations are completely washed out.

\section{Summary and concluding remarks}\label{Sum}

We have discussed the low-density nuclear matter structures,
``nuclear pasta'', and elucidated the charge screening effect.
Using a self-consistent framework based on density functional theory and
relativistic mean fields,
we took into account the Coulomb interaction in a proper way and
numerically solved the coupled equations of motion to extract the
density profiles of nucleons.

First we have checked how realistic our framework is by calculating the bulk
properties of finite nuclei, as well as
the saturation properties of nuclear matter, and found that it can describe
both features satisfactorily.
One could still improve the consideration fitting other experimental data.
For example, we could more carefully fit different terms in the Weiczecker equation
like the surface energy and the shell terms. For that we might be need an
improvement of our relativistic mean field model that does not include the
gradient proton and neutron density terms.

In isospin-asymmetric nuclear matter for fixed proton to baryon number ratios, 
we have observed the ``nuclear pasta''
structures with various geometries at sub-nuclear densities. These cases are
relevant for the discussion of the supernova explosions and for the
description of the newly born neutron stars.
The appearance of the pasta structures significantly lowers the energy,
i.e.\ softens the equation of state, while the energy differences between
various geometrical structures are rather small.
The spatial rearrangement of the proton and electron
charge densities (screening) affect the geometrical structures.

By comparing different treatments of the Coulomb interaction, we
have seen that the self-consistent inclusion of the Coulomb
interaction changes the phase diagram. In particular the region of the
pasta structure is broader for ``full calculation'' compared
to that with simplified treatments of the Coulomb interaction
which have been used in the previous studies. The effect of the
rearrangement of the proton distributions on the structures is
much more pronounced compared to the effect of the electron charge
screening. The influence of the charge screening on the equation
of state, on the other hand, was found to be small.

We have also studied the structure of the nucleon matter in the beta equilibrium.
We have found that  only one type of structures is realized: proton-enriched
droplets embedded in the neutron sea.
No other geometrical structures like rod, slab, etc.\ appeared.

In application to the newly formed neutron stars like in  supernova explosions, finite temperature and
neutrino trapping effects become important, as well as the dynamics of the
first order phase transition with formation of the structures.
It would be interesting to extend our
framework to include these effects.

This work is partially supported by the Grant-in-Aid for the 21st Century COE
``Center for the Diversity and Universality in Physics '' from
the Ministry of Education, Culture, Sports, Science and
Technology of Japan. It is also partially supported by the Japanese
Grant-in-Aid for Scientific
Research Fund of the Ministry of Education, Culture, Sports, Science and
Technology (13640282, 16540246).
The work of D.N.V.\ was also supported in part by the Deutsche
Forschungsgemeinschaft (DFG project 436 RUS 113/558/0-2).

\appendix

\section{``Bulk'' and ``no Coulomb + sharp surface''
calculation of low density nucleon matter}\label{app:bulk}

The bulk calculation proceeds like this \cite{HPS93,pei95,GS99,ARRW,Rav83}:
 first consider two
semi-infinite matters, (I) dense and (II) dilute phases,
with a sharp boundary. The Coulomb and surface interactions
are discarded for a while. 
Conditions of thermal equilibrium at zero temperature are
imposed for pressure $P$ and chemical potential $\mu_a$ ($a=n,p$)
between two phases:
\begin{eqnarray}
P^{\rm (I)}&=&P^{\rm (II)},\nonumber\\
\mu_p^{\rm (I)}&=&\mu_p^{\rm (II)},\nonumber\\
\mu_n^{\rm (I)}&=&\mu_n^{\rm (II)}.\nonumber\\
\label{a1}
\end{eqnarray}
In Sec.~\ref{Ypfix} the case is considered when  the beta
equilibrium is not imposed but the proton to baryon number
ratio $Y_p$ is fixed.  

The averaged densities are $\rho_p=f\rho_p^{(\rm
I)}+(1-f)\rho_p^{(\rm II)}$ and $\rho_n=f\rho_n^{(\rm
I)}+(1-f)\rho_n^{(\rm II)}$. Here $f$ is the volume fraction of
the phase (I). The chemical potentials are calculated using
the RMF model presented in this paper. 
Taking into account the above conditions (\ref{a1}), we obtain 
a set of $\rho_n^{\rm (I)}$, $\rho_n^{\rm
(II)}$, $\rho_p^{\rm (I)}$, $\rho_p^{\rm (II)}$, $P^{\rm
(I)}=P^{\rm (II)}$, $f$, and the bulk energy density
$\epsilon_{\rm bulk}$ in each phase 
for given $\rho_n$ and $\rho_p$. 
At this point $\epsilon_{\rm bulk}$
does not include the surface and Coulomb contributions.
If one cannot find the solution with finite $\rho_n^{\rm (II)}$ and $\rho_p^{\rm (II)}$, 
the proton or neutron  density of the dilute phase
is set to be zero. In this case the corresponding chemical
potential is larger in the phase (II) than in the phase (I), and
the complete set of the Gibbs conditions is not fulfilled.
We leave off here the discussion of the ``bulk calculation''.

Now let us specify the ``no Coulomb + sharp surface''
calculation. To consider the structure of the mixed phase, the
balance between the Coulomb interaction and the surface one
should be taken into account. Introducing an adjusting
parameter of the surface tension $\tau_{\rm surf}$, 
we calculate the surface energy density for the given geometrical dimension $D$:
\begin{eqnarray}
\epsilon_{\rm surf}&=&{\tau_{\rm surf} fD\over R_{d}},
\end{eqnarray}
where  $R_{d}$ is the droplet radius. The Coulomb
energy density can be calculated \cite{Rav83} as
\begin{eqnarray}
&&\epsilon_{\rm Coul}=2\pi e^2\left(\rho_p^{\rm (I)}-\rho_p^{\rm (II)}\right)^2{R_{d}}^2f\Phi,\\
&&\Phi\equiv\left[{2-Df^{1-2/D}\over D-2}+f\right]{1\over D+2}.
\end{eqnarray}
By minimization of  $\epsilon_{\rm surf}+\epsilon_{\rm Coul}$
in $R_{d}$  (the relation $\epsilon_{\rm surf}=2\epsilon_{\rm
Coul}$), we get 
\begin{eqnarray}
R_{d}&=&\left({\tau_{\rm surf} D\over4\pi\left(\rho_p^{\rm (I)}-\rho_p^{\rm (II)}\right)^2\Phi}\right)^{1/3},\\
\epsilon_{\rm Coul}+\epsilon_{\rm surf}&=&3fD\left({\pi\tau_{\rm surf}^2
\left(\rho_p^{\rm (I)}-\rho_p^{\rm (II)}\right)^2\Phi\over 2D}\right)^{1/3}.
\end{eqnarray}
Comparing the energy density of the uniform matter
$\epsilon$ and those of mixed phases $\epsilon_{\rm
bulk}+\epsilon_{\rm surf}+\epsilon_{\rm Coul}$ with different
geometrical dimension $D$, we can determine the most favorable
configuration and its energy density.


\begin{thebibliography}{99}
\bibitem{gle92}
N.~K.~Glendenning, Phys. Rev. {\bf D46}, 1274 (1992);
N.~K.~Glendenning, Phys. Rep. {\bf 342}, 393 (2001).

\bibitem{HPS93}
H.~Heiselberg, C.~J.~Pethick and E.~F.~Staubo, Phys. Rev. Lett. {\bf
70}, 1355 (1993).

\bibitem{pei95}
N.~K.~Glendenning and S.~Pei, Phys. Rev. C {\bf D52}, 2250 (1995).

\bibitem{voskre}
D.~N.~Voskresensky, M.~Yasuhira and T.~Tatsumi,
Phys. Lett. {\bf B541}, 93 (2002);  T.~Tatsumi and D.~N.~Voskresensky, nucl-th/0312114.

\bibitem{voskre1}
D.~N.~Voskresensky, M.~Yasuhira and T.~Tatsumi, Nucl. Phys.
{\bf A723}, 291 (2003).

\bibitem{emaru1}
T. Endo, Toshiki Maruyama, S. Chiba and T. Tatsumi, nucl-th/0410102; 
hep-ph/0502216. 

\bibitem{GS99}
N.~K.~Glendenning and J.~Schaffner-Bielich, Phys. Rev. {\bf C60}, 
025803 (1999).

\bibitem{CG00}
M.~B.~Christiansen and N.~K.~Glendenning, astro-ph/0008207.

\bibitem{CGS00}
M.~Christiansen, N.~K.~Glendenning and J.~Schaffner-Bielich, Phys.
Rev. {\bf C62}, 025804 (2000).

\bibitem{PREPL00}
J.~A.~Pons, S.~Reddy, P.~J.~Ellis, M.~Prakash and  J.~M.~Lattimer,
Phys. Rev. {\bf C62}, 035803 (2000).

\bibitem{MYTT}
M.~Yasuhira and T.~Tatsumi, Nucl. Phys. {\bf A690}, 769 (2001).

\bibitem{RB}
 S. Reddy, G. Bertsch and M. Prakash, Phys. Lett. {\bf B475}, 1 (2000). 

\bibitem{NR00}
T.~Norsen and S.~Reddy, Phys. Rev. {\bf C63}, 065804 (2001).


\bibitem{maruKaon}
Toshiki ~Maruyama, T.~Tatsumi, D.~N.~Voskresensky, T.~Tanigawa and S.~Chiba,
 Nucl. Phys. {\bf A749}, 186 (2005).

\bibitem{dubna}
T.~Tatsumi, Toshiki ~Maruyama, D.~N.~Voskresensky, T.~Tanigawa and
	S.~Chiba, nucl-th/0502040.


\bibitem{ARRW}
M.~Alford, K.~Rajagopal, S.~Reddy and F.~Wilczek, Phys. Rev.
{\bf D64}, 074017 (2001).

\bibitem{bed}
P.~F.~Bedaque, Nucl. Phys. {\bf A697}, 569 (2002).

\bibitem{RR04}
S.~Reddy and G.~Rupak, nucl-th/0405054.



\bibitem{BCR03}
P.~F.~Bedaque, H.~Caldas and G.~Rupak, Phys. Rev. Lett. {\bf 91}, 247002 (2003).




\bibitem{Rav83} 
D.~G.~Ravenhall, C.~J.~Pethick and J.~R.~Wilson,
Phys. Rev. Lett. {\bf 50}, 2066 (1983).


\bibitem{Has84} 
M.~Hashimoto, H.~Seki and M.~Yamada,
Prog. Theor. Phys. {\bf 71}, 320 (1984).

\bibitem{Wil85} 
R.~D.~Williams and S.~E.~Koonin,
Nucl. Phys. {\bf A435}, 844 (1985).

\bibitem{Oya93} 
K.~Oyamatsu,
Nucl. Phys. {\bf A561}, 431 (1993).

\bibitem{Lor93} 
C.~P.~Lorenz, D.~G.~Ravenhall and C.~J.~Pethick,
Phys. Rev. Lett. {\bf 70}, 379 (1993).

\bibitem{Cheng97}
K.~S.~Cheng, C.~C.~Yao and Z.~G.~Dai,
Phys. Rev. {\bf C55}, 2092 (1997).

\bibitem{Mar98}
Toshiki~Maruyama, K.~Niita, K.~Oyamatsu, Tomoyuki ~Maruyama, S.~Chiba and A.~Iwamoto,
Phys. Rev.  {\bf C57}, 655 (1998).

\bibitem{kido00}
T.~Kido, Toshiki~Maruyama, K.~Niita and S.~Chiba,
Nucl. Phys. {\bf A663}-{\bf 664}, 877 (2000).

\bibitem{Gen00}
G.~Watanabe, K.~Iida and K.~Sato,
Nucl. Phys. {\bf A676}, 445 (2000);

\bibitem{Gen02}
G.~Watanabe, K.~Sato, K.~Yasuoka and T.~Ebisuzaki,
Phys. Rev.  {\bf C66}, 012801(R) (2002).

\bibitem{Gen03}
G.~Watanabe and K.~Iida,
Phys. Rev.  {\bf C68}, 045801 (2003).

\bibitem{horowitz} 
C.~J.~Horowitz, M.~A.~P\'erez-Garc\'ia, and J.~Piekarewicz,
Phys. Rev.  {\bf C69}, 045804 (2004).

\bibitem{MI95} 
Y.~Mochizuki and T.~Izuyama,
Astrophys. J. {\bf 440}, 263 (1995).

\bibitem{maru1}
Toshiki~Maruyama, T.~Tatsumi, D.~N.~Voskresensky, T.~Tanigawa,  S.~Chiba and
Tomoyuki Maruyama, nucl-th/0402202.


\bibitem{refDFT}
{\it Density Functional Theory}, ed. E.~K.~U.~Gross and R.~M.~Dreizler,
Plenum Press (1995).


\bibitem{voskresurf}
D.~N.~Voskresensky et al., in preparation.

\bibitem{centelles93}
M. Centelles and X. Vi\~nas,
Nucl. Phys. {\bf A563}, 173 (1993).







\end{thebibliography}
\end{document}